\documentclass[twocolumn]{aastex631}
\let\tablenum\relax
\usepackage{siunitx}

\usepackage{cleveref}

\begin{document}

\title{Gaia-GIC-1: An Evolving Catastrophic Planetesimal Collision Candidate}

\author[0000-0003-0484-3331]{Anastasios Tzanidakis}\thanks{Corresponding Author: \href{mailto:atzanida@uw.edu}{atzanida@uw.edu}}
\affiliation{Department of Astronomy and the DiRAC Institute, University of Washington, 3910 15th Avenue NE, Seattle, WA 98195, USA}

\author[0000-0002-0637-835X]{James R. A. Davenport}
\affiliation{Department of Astronomy and the DiRAC Institute, University of Washington, 3910 15th Avenue NE, Seattle, WA 98195, USA}

\begin{abstract}
We report the discovery of the optical dipper and low-luminosity infrared stellar transient Gaia20ehk (hereafter, Gaia-GIC-1), which is currently undergoing high-amplitude variability due to transiting dusty material. In this work, we identify Gaia-GIC-1 as a likely young F-type star based on the spectral energy distribution before the onset of the high-amplitude optical variability.  We detect a significant periodic modulation of 380.5 days in Gaia-G band before the onset of the infrared brightening, consistent with a $\sim$1.1 AU orbit assuming circular orbits and a 1.3 M$_{\odot}$ star. The system has remained in an infrared bright state for $>$4 years since the last near-infrared detection, confirmed by recent SPHEREx observations, while continuing to undergo large amplitude irregular optical dimming. We measure the dust temperature from the freshly generated debris to be $\sim$900 Kelvin based on available WISE photometry, and the dust clump size to have a minimum cross-sectional area of 0.13 AU$^{2}$, and the dust mass $\SI{4e20}{\kilogram}$. Currently, optical follow-up spectroscopy has not revealed any prominent features in the system, likely due to its highly variable nature. We hypothesize that Gaia-GIC-1 represents debris recently formed in a planetary collision, which produced a clumpy dust cloud on a bound orbit, producing the observed dimming events. The ongoing collisional activity in this system presents a unique opportunity for understanding terrestrial planet formation.
\end{abstract}


\section{Introduction} \label{sec:intro}

Giant impacts represent some of the most fundamental events in late-stage planetary formation. Canonical planet formation theories predict that terrestrial planets reach their final mass assembly through collisions between up to Mars-sized planetesimals, typically during the first $\sim$100 Myr of the system evolution \citep[][and references therein]{Wyatt_Jackson_Review}. Giant impact collisions are expected to be a common occurrence during the terrestrial planet formation era \citep{Canup_GI}, with collision-generated debris disks persisting for millions of years and producing detectable infrared (IR) excess signatures \citep{Wyatt_2008}. These collisions are thought to be a primary driver of warm debris disks and occur 10\% around young stars \citep{Genda_2015, 2012MNRAS.425..657J}. Numerical simulations indicate that most Earth-like planets experience several giant impact collision events within 2 Gyr, with the vast majority occurring in the first 200 Myr \citep{Quintana_GI_freq}. However, these simulations are computationally intensive, and a new generation of machine learning accelerated simulations is being developed to explore wider ranges of impact velocities and masses \citep{LammersGI_ML}. 

Time-domain monitoring is now surfacing plausible giant-impact aftermaths among young stars. Intensive photometric monitoring of extreme debris disks (EEDs) reveals week-to-month-long mid-IR and optical modulations that are consistent with optically thick post-impact clumps that shear around the orbits and then collisionally grind down \citep{Su_2019, Moor_TYC_4209, Su_HD166191, Reike_V488, Kenworthy_ASASSN21qj}. Within favorable viewing geometries and impact conditions, excess IR photometric variability is directly connected to the reprocessing of heat and dissipation of freshly generated debris \citep{Meng2014, Su_HD166191, Reike_V488}, while the irregular optical variability arises due to the occultation of dusty sheared clumps \citep{Jackson2014_planet_embroys}. Optical monitoring of young main-sequence stars has also revealed complex irregular dips with timescales spanning days to years, which have often been attributed to the impact collisions of planetary embryos, though with little presence of IR excess \citep{Boyajian_KIC2852, Tzan2025, Zakamska_ASASSN24fw, big_dippers_assassn, deWit_RZ}. The highly variable nature of impact collisions contradicts the scenario of steady-state cascades, and instead points to violent collisions among large bodies in the inner AU. 

In this study, we present the discovery of Gaia-GIC-1, a low-luminosity infrared transient with simultaneous optical dimming, resulting from what we believe to be the collision of large planetesimals. This is the first such system discovered with the Gaia Alerts, and one of a very small number of such giant impact candidates known. Section \ref{sec:data} presents the details of the discovery and follow-up photometric and spectroscopic data. In Section \ref{sec:properties}, we lay out the light curve properties of the discovered system. In Section \ref{sec:discussion}, we provide a discussion on the mass and properties of the occulting dust clump. We summarize our conclusions in Section \ref{sec:conclusion} and outline future observations that could further illuminate the nature of this and similar systems.

\section{Data} \label{sec:data}

\begin{figure*}
\centering
\includegraphics[width=0.75\linewidth]{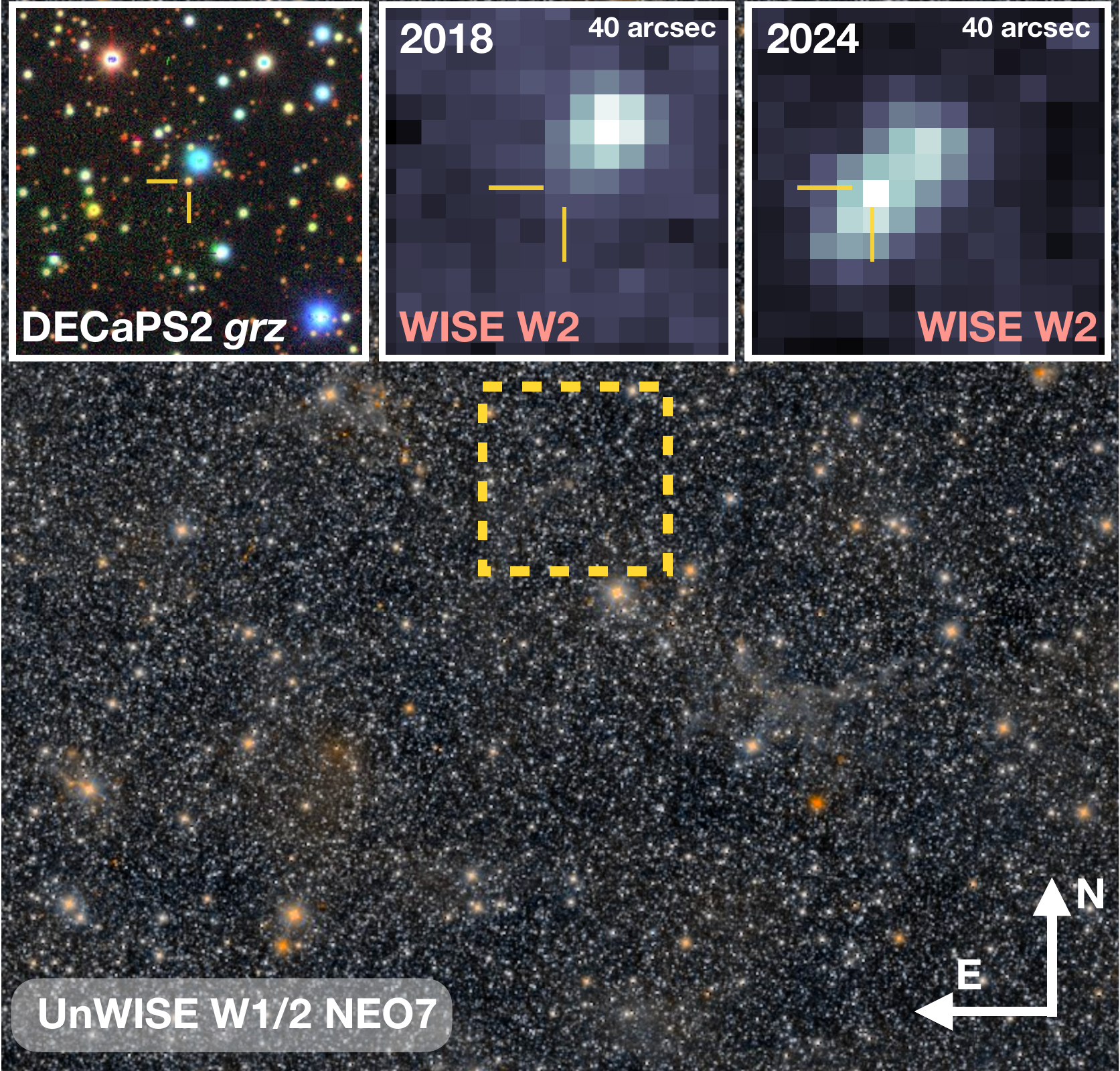}
\caption{Multi–wavelength view of Gaia-GIC-1. \textbf{Top‑left:} DECaPS2 three‑color ($g\,r\,z$) mosaic, $40\arcsec\times40\arcsec$, with the identified star marked by yellow cross‑hairs. \textbf{Top‑middle} and \textbf{top‑right:} WISE/NEOWISE $W2$ ($4.6\,\mu$m) cut‑outs of identical size from 2018 and 2024, respectively, showing an increase in mid‑IR brightening and a subtle centroid shift relative to the neighboring field. \textbf{Bottom:} unWISE $W1/W2$ NEO7 coadd. The dashed yellow box indicates a 20$'$x20$'$ region around Gaia-GIC-1.}
\label{fig:skypos}
\end{figure*} 

\subsection{Optical Light Curves} \label{sec:optical}

\begin{figure*}
    \centering
\includegraphics[width=0.9\linewidth]{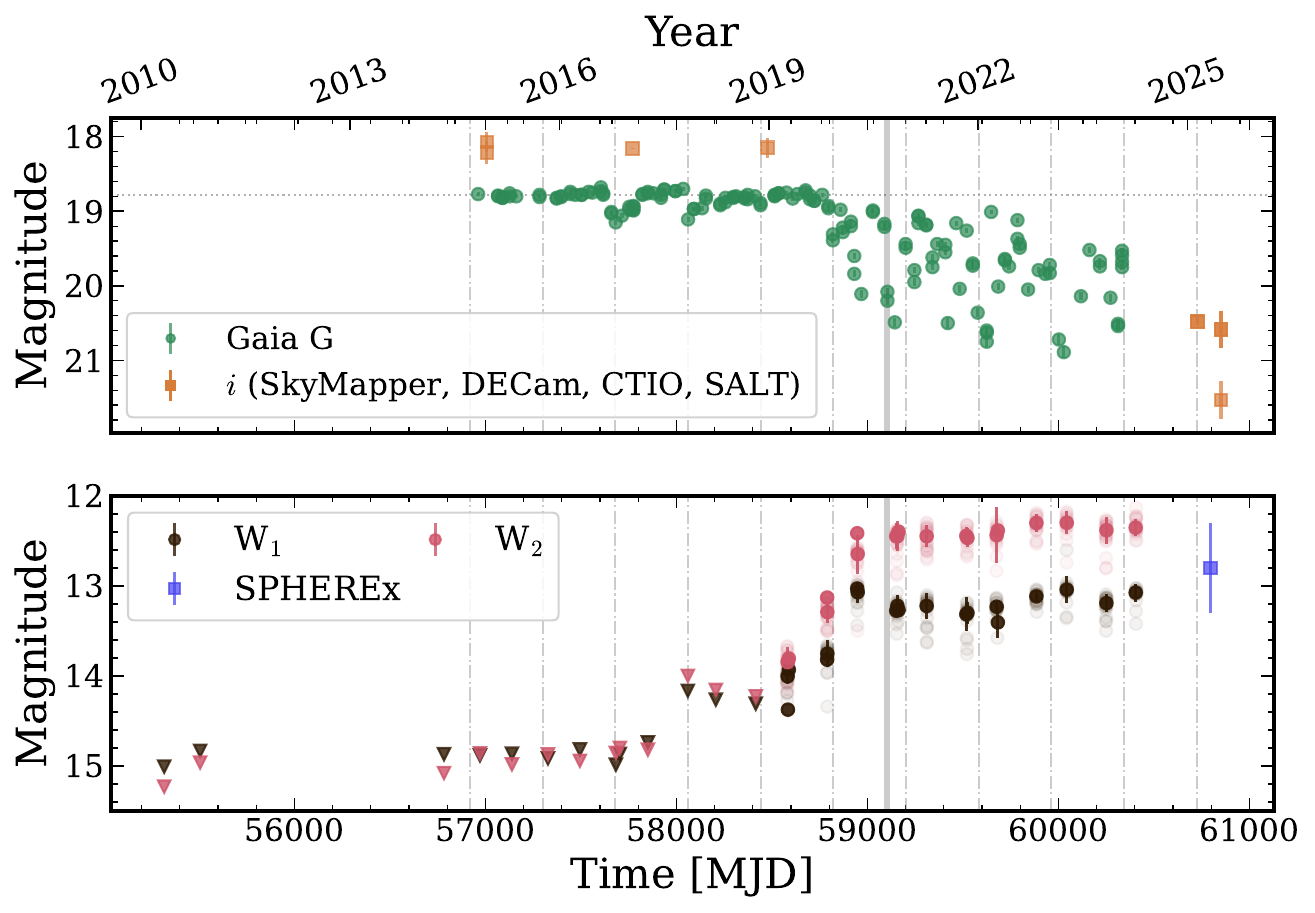}
    \caption{Compiled optical and infrared light curve of Gaia-GIC-1. (\textbf{Top}) Optical Gaia-$G$ and $i$ band light curves. The dashed horizontal line is the median Gaia-$G$ magnitude in quiescence. (\textbf{Bottom}) WISE photometry in the W$_{1}$ and W$_{2}$, where upside-down triangles indicate the 3-$\sigma$ upper limits. The dashed vertical gray lines mark a 380.5-day interval, and the highlighted solid gray line is the epoch of the Gaia alert. The data are available as a machine-readable table.}
    \label{fig:gaia_wise_lc}
\end{figure*}

Gaia20ehk (AT~2020tdg; \citealt{atel}) was first identified by the \textit{Gaia} Photometric Science Alerts (GPSA) \citep{GaiaAlerts_Hodgkin}. We retrieved the GPSA G-band light curve with \texttt{GaiaAlertsPy}\footnote{\href{https://github.com/AndyTza/GaiaAlertsPy}{\texttt{github.com/AndyTza/GaiaAlertsPy}}} \citep{2023ApJ...955...69T}. Archival $i$-band detections were gathered from SkyMapper DR4 \citep{skymap_dr4} and DECam Plane Survey \citep{decaps} (DECAPS). Additional follow-up $i$-band proprietary photometry was obtained on behalf of the Microlensing Telescope Network (KMTNet; \citet{KMTNET}) from the Cerro-Tololo Inter-American Observatory (CTIO) and the South African Astronomical Observatory (SAAO) sites. One $i$-band detection was also provided through the SALTICAM target acquisition during spectroscopic follow-up, discussed in Section \ref{sec:spec}. Standard aperture photometry was performed on the obtained images using $\texttt{photutils}$\footnote{\href{https://photutils.readthedocs.io/}{photutils.readthedocs.io/}}. We present an image of the sky location mosaic of Gaia-GIC-1 in Figure \ref{fig:skypos} identified by several surveys. The collated optical light curve is shown in the top panel of Figure~\ref{fig:gaia_wise_lc}.

\subsection{Infrared Light Curve} \label{sec:IR}

Initial epochal infrared data were obtained using WISE unTIMELY \citep{unTimely}, which contained epochs of data up until December 2020 that include up to year 7 of NEOWISE data. To obtain WISE data after 2020, we used the NASA IPAC Infrared Science Archive (IRSA) to download the WISE single epoch images in the W$_{1}$ and W$_{2}$ bandpasses of all WISE scans between 2021 and 2024 \citep{WISE_L1b}. We ran forced photometry on the Gaia sky position of Gaia-GIC-1 with a general 2D Gaussian PSF model from $\texttt{photutils}$. We converted our PSF flux to Vega magnitudes using zero-point offsets \citep{Wright_WISE}. Overall, we find that our derived forced PSF photometry measurements are consistent with the magnitude measurements made using the unTIMELY light curve. Before the onset of the IR outburst, we are unable to draw robust detections of Gaia-GIC-1 in the W$_{1}$ and W$_{2}$, respectively, due to its faintness and proximity to a nearby source. We applied a customized image subtraction pipeline based on the ZOGY algorithm \citep{De_2020, Zackay_2016} on the unWISE images. The resulting difference images yielded no detections above a 5$\sigma$ threshold prior to MJD 58,500 in either the W$_{1}$ or W$_{2}$ bands. We therefore treat all WISE photometry before MJD 58,500 as upper limits in our subsequent analysis, demonstrated in Figure \ref{fig:gaia_wise_lc}. We searched through the archival Infrared Astronomical Satellite (IRAS) images and did not find any corresponding source at the location of Gaia-GIC-1, except for one source $\sim 0.5$ deg away, which is unlikely to be associated with Gaia-GIC-1. The most recent detection between 1-4.5 $\mu$m was made by the SPHEREx mission \citep{Spherex_Dore}. Flux measurements were obtained via the IRSA SPHEREx spectrophotometry tool. We convolved the linearly interpolated SPHEREx spectrum with the NEOWISE W$_{1}$ and W$_{2}$ bandpasses using $\texttt{pyphot}$ \citep{zenodopyphot} and averaged the resulting Vega magnitudes. We note that photometry for this source with SPHEREx is challenging due to the large pixel-scale size of the SPHEREx detector (6.2 arcseconds), and with the nearest stellar source 8.2 arcseconds away. While the SPHEREx detection makes it clear that Gaia-GIC-1 remains IR bright, additional follow-up photometry is needed to monitor its brightness in the optical and infrared.

\subsection{Spectroscopy} \label{sec:spec}

Optical spectroscopy of Gaia-GIC-1 was obtained using the Goodman High Throughput Spectrograph on the 4.1-meter Southern Astrophysical Research (SOAR) telescope. The initial spectrum was acquired on 2024 December 19 as part of the SOAR2024B-026 program. Observations were conducted using the Red Camera with the 400 lines/mm grating (centered at approximately 600 nm), providing a spectral resolution of $R\approx2000$. The SOAR reductions were done using $\texttt{PypeIt}$ \citep{pypiet}. Additional spectroscopic observations were obtained with the Robert Stobie Spectrograph (RSS) on the Southern African Large Telescope (SALT) on 2025 February 22 through a Director's Discretionary Time. Observations utilized the Pg0900 grating with a 1-arcsecond slit in ``Faint'' gain and ``slow'' readout modes, yielding a spectral resolution of $R \approx 1700$-$2300$. Two consecutive 20-minute exposures were acquired to allow for cosmic ray rejection. One acquisition image was taken with the SALTCAM finder in the $i$-band. Spectroscopic reductions were performed using the RSS reduction pipeline\footnote{\href{https://astronomers.salt.ac.za/software/rss-pipeline/}{https://astronomers.salt.ac.za/software/rss-pipeline/}}, which applies bad pixel replacement, cosmic ray cleaning, CCD gap filling, wavelength calibration, auto gain correction, and flat fielding.

Both optical spectra of Gaia-GIC-1 have a low signal-to-noise ratio (SNR), and are thus difficult to extrapolate any meaningful constraints from either the primary star or any occulting material, likely due to the ongoing large dimming variability Gaia-GIC-1 is experiencing. We fit and divide out the local continuum using a line-free window around the H$\alpha$ line within $\pm \text{10 Å}$, and integrate the continuum-normalized spectrum to measure the equivalent-width (EW) and estimate the 1$\sigma$ EW uncertainty by propagating the scatter from the normalized continuum. Both SALT and SOAR did not meet our 5$\sigma$ threshold criterion, and we thus report the upper limits of the $|\mathrm{EW}_{H_\alpha}| <10.3 \text{Å}$ and $|\mathrm{EW}_{H_\alpha}| <3 \text{Å}$ for SALT and SOAR, respectively. In the SALT spectrum, we identified weak features of the Ca II infrared triplet lines $\lambda \lambda$8498, 8542, and 8662 Å, which would be consistent with photospheric absorption features seen in FGK dwarfs \citep{Chmielewski_CaIIRT}.

\begin{figure}
    \centering
    \includegraphics[width=1.0\linewidth]{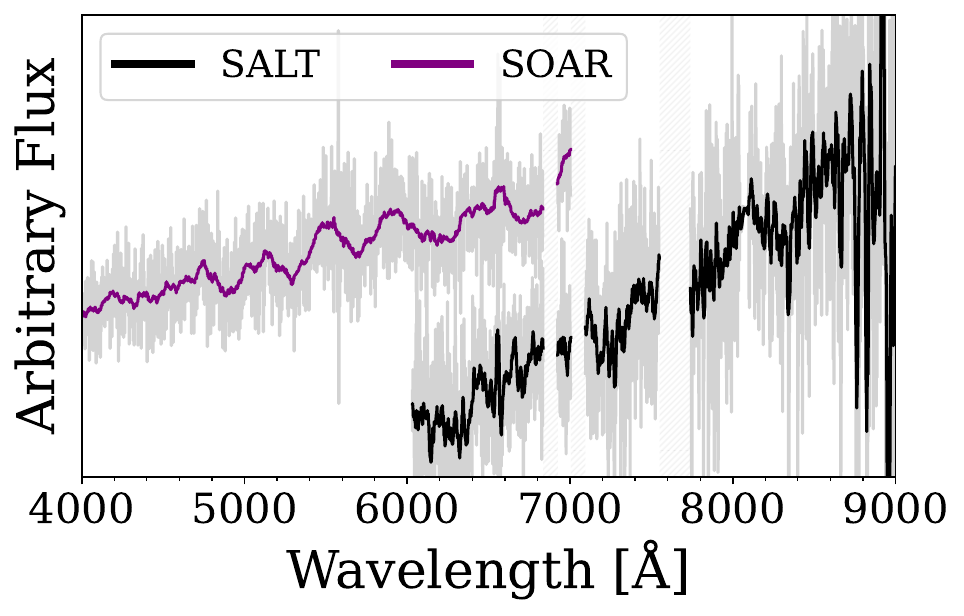}
    \caption{Optical spectra of Gaia-GIC-1 obtained with SOAR (purple) and SALT (black) telescopes. Gray lines show the unbinned spectra, while the colored lines show binned versions for clarity. Flux is given in arbitrary units.}
    \label{fig:periodogram}
\end{figure} 

\subsection{Spectral Energy Distribution} \label{sec:sed_fit_section}

\begin{figure*}
    \centering \includegraphics[width=0.65\linewidth]{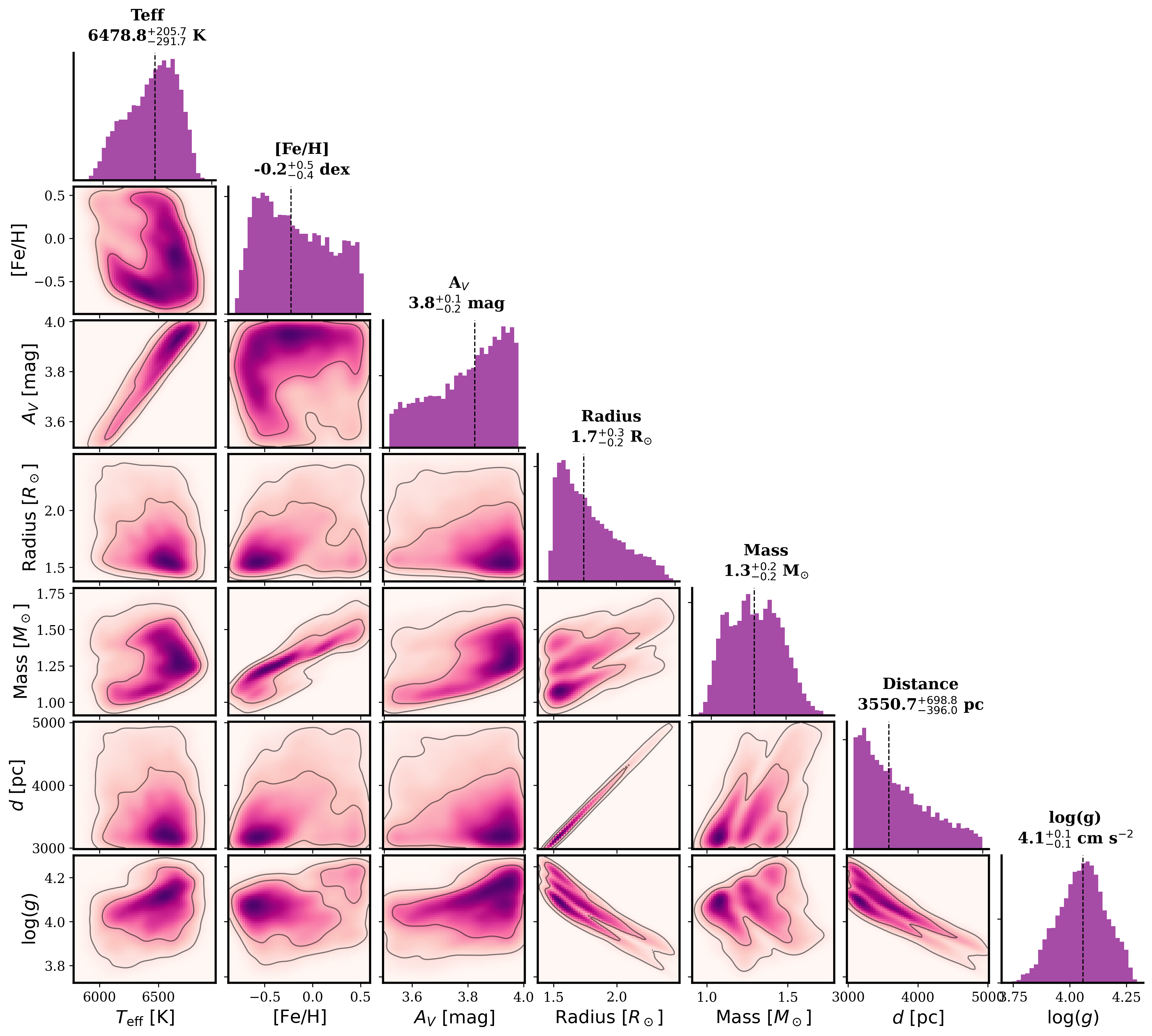}
    \caption{Corner plot showing posterior distributions from MIST fitting to pre-eclipse photometry of Gaia-GIC-1. Diagonal panels show marginalized posteriors for Teff, [Fe/H], $A_V$, radius, mass, distance, and log g. The off-diagonal panels show parameter correlations with 68\% and 95\% confidence contours.}
    \label{fig:SED_fit}
\end{figure*}

We constructed the broadband spectral energy distribution (SED) using available archival data of the star before the onset of the IR brightening event to avoid any contamination from excess IR emission. Our SED compilation incorporates photometric measurements from Gaia BP/G/RP photometry \citep{2023A&A...674A...1G} and 2MASS $HJK_{s}$ bands \citep{2mass} presented in Table \ref{table:pre-eclipse-photometry}. To perform the SED fitting, we used the Modules for Experiments in Stellar Astrophysics \citep[MESA][]{Paxton2011, Paxton2013, Paxton2015, Paxton2018, Paxton2019} Isochrones \& Stellar Tracks using the $\texttt{isochrones}$ package \citep{2015ascl.soft03010M} to generate synthetic photometry while simultaneously fitting for fundamental stellar parameters, distance, and extinction presented in Figure \ref{fig:SED_fit} using $\texttt{MultiNest}$ \citep{2009MNRAS.398.1601F}. We adopt broad uniform priors for all stellar parameters, uniform 16-84$^{th}$ percentile photogeometric distance from \citet{BailerJonesDR3}, and uniform extinction values from 1-4 mag based on the E(B-V) values reported by \citet{SFD_dust}. Within 30 arc-minutes of Gaia-GIC-1, we identified two open clusters, FSR 1347 and FSR 1352 \citep{Hunt_2024}. Though Gaia-GIC-1 is not flagged as a member of either cluster, both clusters have derived isochronal ages between 6-16 Myr, extinctions A$_{V}$ 3-3.9 mag, and distances 3.29 - 3.38 kpc, consistent within the posterior distribution margin derived for Gaia-GIC-1 from \citet{BailerJonesDR3}. The astrometric properties of Gaia-GIC-1 are broadly consistent with the FSR 1347/1352 region, with marginal evidence suggesting a possible cluster membership association with FSR 1352 (see \hyperref[ap:apendixA]{Appendix~\ref*{ap:apendixA}}).

To determine whether the available photometry constrains the stellar age, we performed two SED fits using different age priors. First, motivated by the potential association with FSR 1347/1352, we adopted a narrow uniform age prior $\log_{10}(t/\mathrm{yr}) \in [6.8,\,7.5]$ based on the clusters' ages, while maintaining our broad distance and A$_{V}$ priors. Next, we ran a second iteration with a broad age prior $\log_{10}(t/\mathrm{yr}) \in [6.8,\,9.5]$ under the scenario that Gaia-GIC-1 is a field star and test if the posterior distribution favors a young age solution. In our broad age prior model, the posterior distribution peaks with a $\log_{10}(t/\mathrm{yr})\sim 9^{+9.2}_{-8.1}$, while the narrow age model $\log_{10}(t/\mathrm{yr})\sim 7.1^{+7.2}_{-6.8}$. Comparing our SED fits, we measured a Bayes factor of $\sim 1$, indicating no preference for either model. While the stellar age cannot be reliably constrained using the available photometry and MIST isochrones alone, the derived stellar parameters remained consistent between both fits. From our fit presented in Figure~\ref{fig:SED_fit}, we derive a stellar effective temperature of $T_{\mathrm{eff}} = 6478.8^{+206}_{-292}$~K, metallicity of $[\mathrm{Fe/H}] = -0.2^{+0.5}_{-0.4}$ dex, and extinction of $A_{V} = 3.8^{+0.1}_{-0.2}$~mag. From this fit, we assume a radius of $R_{\star} = 1.7^{+0.3}_{-0.2} ~R_{\odot}$, mass of $M_{\star} = 1.3^{+0.2}_{-0.2} ~M_{\odot}$, distance $d = 3551^{+699}_{-396}$~pc, surface gravity $\log g = 4.1^{+0.1}_{-0.1}$ cgs (\dataset[10.5281/zenodo.17982426]{https://doi.org/10.5281/zenodo.17982426}). Based on the SED posteriors, we interpret Gaia-GIC-1 as a mid-F5 star.

\begin{table}[ht]
\centering 
\begin{tabular}{c c c} 
\hline\hline 
Survey/Band & Magnitude & Magnitude Error\\ [0.05ex]
\hline 
2MASS-H                         & 15.45 & 0.02 \\
2MASS-J                         & 16.19 & 0.09 \\
2MASS-Ks\footnote{Possible spurious detection with SNR$\sim$5} 
                                & 15.41 & 0.21 \\
\hline
Gaia DR2 - BP                          & 19.68 & 0.06 \\
Gaia DR2 - G                           & 18.79 & 0.002 \\
Gaia DR2 - RP                           & 17.71 & 0.02 \\
\hline \hline 
\end{tabular}
\caption{Compiled broadband median photometry of Gaia-GIC-1 outside the dimming events before December 2011 or after January 2019. We assume that the intrinsic stellar SED is unchanged before or after the dimming event.}
\label{table:pre-eclipse-photometry}
\end{table}

\section{Properties} \label{sec:properties}

\subsection{Light Curve Properties}

From the beginning of the \textit{Gaia} coverage until approximately ${\rm MJD}\lesssim58\,300$ (roughly 2014–2017), the source remained in a quiescent state at $G\approx18.8$\ mag. At least three distinct optical dips are present in Figure \ref{fig:gaia_wise_lc} between epochs ${\rm MJD}\sim57\,000$ and ${\rm MJD}\sim58\,800$ with average depth of $\sim$0.4 mag (25$\%$ reduction in flux) and average timescale of 200 days, though we note that curiously the third dip at epoch $\sim$58500 was not as deep as the previous dips. While sparse, archival $i$-band data from DECaPS and SkyMapper DR4 captured one epoch during the first optical dimming event. Notably, during the first dimming event, the Gaia-G band dropped by 0.4 mag, while the $i$-band remained nearly constant, producing a $\Delta$(G-$i$)$\sim$0.8 mag, which could hint that the obscuring material may be wavelength-dependent. Starting near ${\rm MJD}\sim58\,700$, the system underwent rapid fading and variability that has been ongoing for at least 4 years, reaching $G\approx20.3$\ mag just before the automated GPSA alert trigger. The average fading rate during this interval is $\sim0.4$\ mag yr$^{-1}$. Following the alert epoch (${\rm MJD}\gtrsim59\,000$; late 2020 onward), the star has remained faint in optical wavelengths, fluctuating between $G\approx19.8$ and at least $>20.8$ mag. Unlike the earlier monotonic behavior, the post-alert light curve is dominated by irregular variations, with no sign of a sustained re-brightening trend across the current Gaia baseline; however, higher-cadence optical monitoring is required to confirm. 

\begin{figure}
    \centering
    \plotone{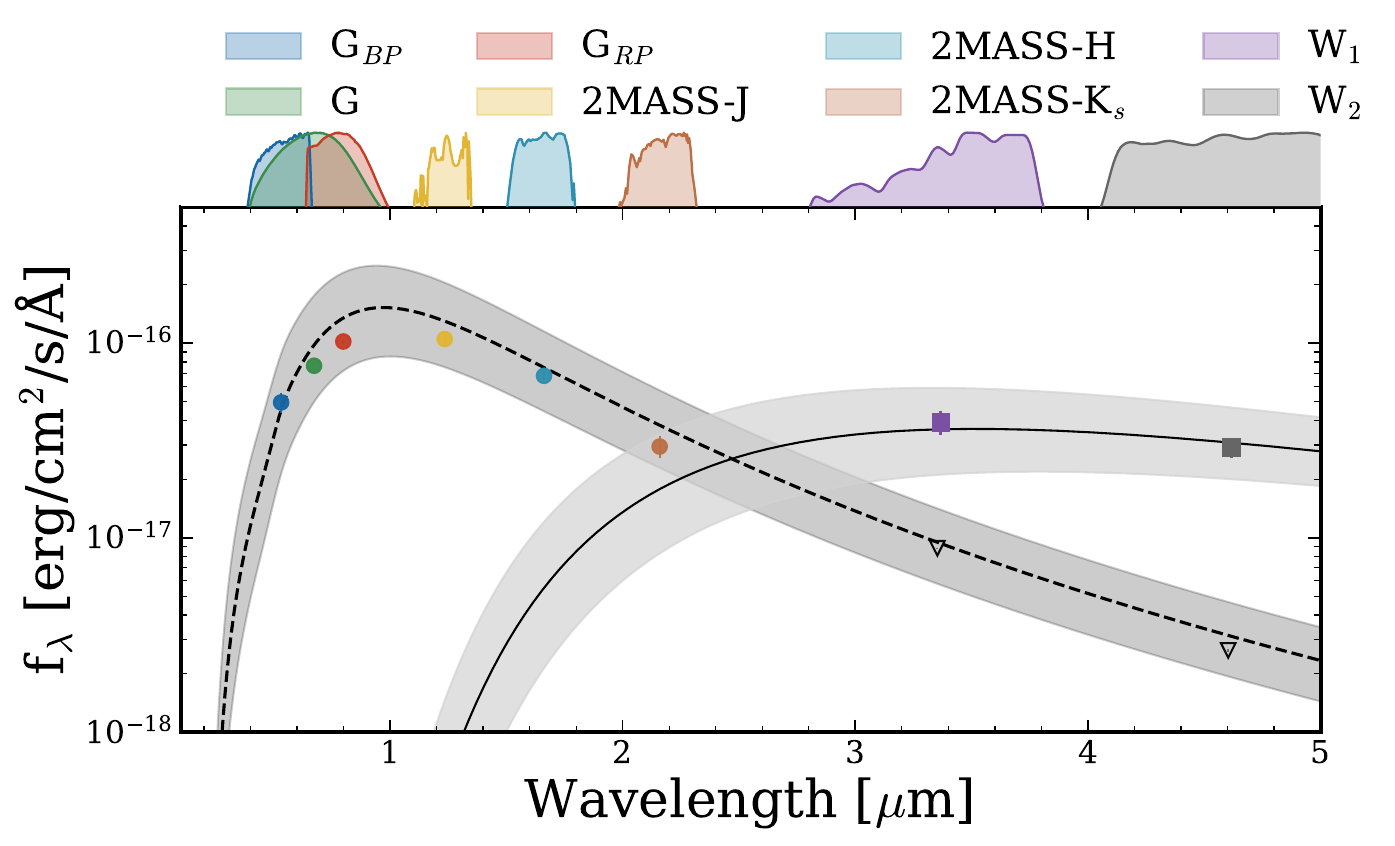}
    \vspace{0.15in}
    \plotone{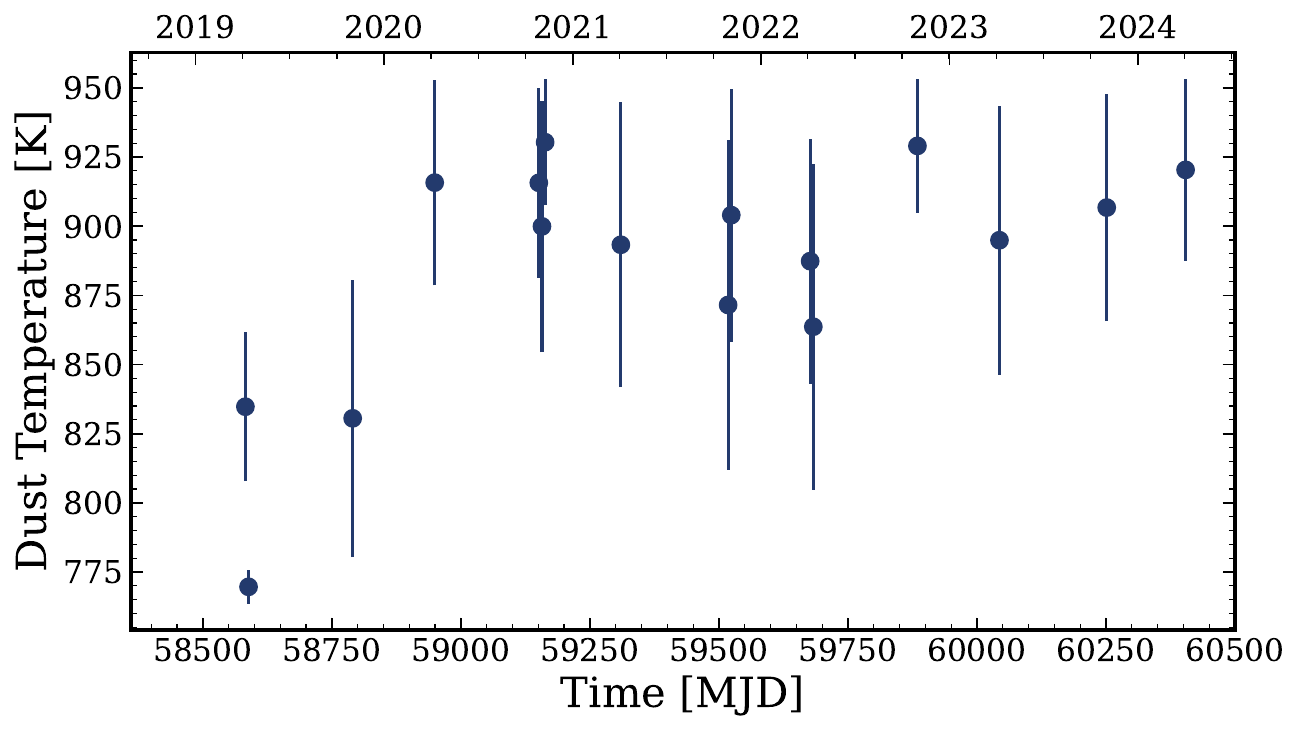}
    \caption{
    (\textbf{Top}) Spectral energy distribution of Gaia-GIC-1 spanning optical to mid-infrared wavelengths. The black dashed line shows the stellar photosphere model derived from MIST isochrone fitting, with the light gray shaded region indicating the associated uncertainty, and the upside-down triangles denote the $3\sigma$ upper limits. The solid black line represents a modified blackbody fit to the infrared excess emission after subtracting the photospheric contribution. (\textbf{Bottom}) Dust temperature evolution after subtracting the stellar photospheric contribution. The data are available as a machine-readable table.}
    \label{fig:bbfit_and_tdustevol}
\end{figure}

The optical evolution is strikingly anti‑correlated with the IR behavior displayed in the bottom panel of Figure \ref{fig:gaia_wise_lc}, where W$_1$ and W$_2$ brighten while the Gaia-$G$ magnitude fades. However, the IR flux has exhibited a characteristic plateau since 2019. The recent SPHEREx detection has also confirmed that Gaia-GIC-1 is currently bright in the IR. Optical dimming coupled with infrared brightening typically indicates newly formed circumstellar dust obscuring reprocessed thermal IR emission. To measure the dust temperature, we fit a modified blackbody to the epochal WISE photometry after subtracting the stellar contribution from the stellar parameters derived in Section~\ref{sec:sed_fit_section}. We use the following modified blackbody: 
\begin{equation}
S_{\lambda} \propto \bigg{(}\frac{\lambda_0}{\lambda}\bigg{)}^{\beta} B_{\lambda}(T_{d}),
\end{equation}
where $\lambda_0$ is the wavelength at which the optical depth is unity, and $\beta$ is the spectral emissivity index. Since far-IR observations of Gaia-GIC-1 are currently unavailable to measure the emissivity index directly, we adopt a $\beta$ value of 0.9 based on the grain-size distribution index in debris disks undergoing collisional cascades \citep{Steele_2016}, and set $\lambda_0$=100$\mu$m. We simultaneously constrain the dust temperature and radius using MCMC. Figure \ref{fig:bbfit_and_tdustevol} shows the blackbody fits (top) and dust temperature evolution (bottom). We measure the median posterior distribution dust temperature, which has plateaued at T$_{dust}$=900$\pm$24~K with a characteristic radius of 0.2$\pm$0.04 AU.

\subsection{Periodicity}

\begin{figure*}
    \centering
    \includegraphics[width=0.95\linewidth]{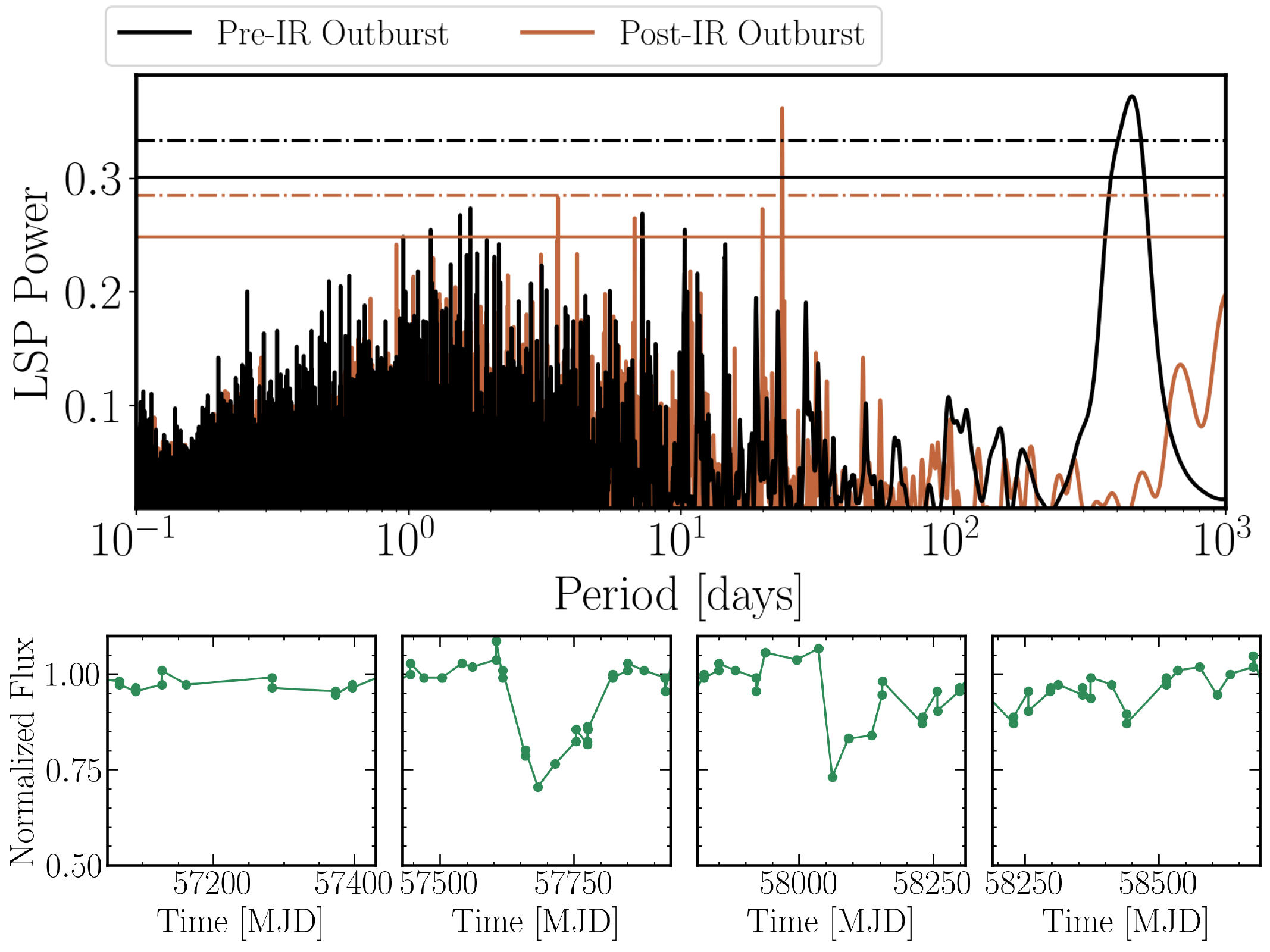}
    \caption{Top panel displays the Lomb-Scargle Periodogram of the optical Gaia light curve of Gaia-GIC-1. The dashed and solid lines, bottom to top, are the 5$\%$ and 1$\%$ False Alarm Probability levels. The bottom panel displays the normalized flux light curve of Gaia-GIC-1 at a 380.5-day interval.}
    \label{fig:periodogram}
\end{figure*} 

We computed the Lomb-Scargle Periodogram \citep{lomb, scargle} on the Gaia-G light curve, before and after the onset of the near-IR excess, separated at the epoch of 2020-09-12, shown in Figure \ref{fig:periodogram}. We find a significant periodic signal at the 380.5-day interval above the 1$\%$ False Alarm Probability (FAP) interval, which diminishes during the IR bright phase, and is consistent with the optical Gaia-G light curve in the bottom panel of Figure \ref{fig:periodogram}. The FAP intervals were measured using 1,000 bootstrap samples with replacement on the flux values while keeping the observed times. Using our estimated stellar mass of 1.3~M$_{\odot}$ and the measured 380.5-day orbital period, we infer that the dusty material is at a semi-major axis of 1.1 AU. At ${\rm MJD}\sim58\,790$ with the onset of the IR-excess feature, the Gaia-G light curve shows more erratic variability, with no clear signs of periodicity. To test this further, we computed the quasi-periodicity metric (Q) \citep{Cody14, Cody18} based on the top 5 highest peaks in the post-IR periodogram in Figure \ref{fig:periodogram} and in all tested periods, we found high Q$>$0.8 values, indicating that the source is likely experiencing aperiodic behavior during the IR-bright phase.  

\section{Discussion} \label{sec:discussion}

\subsection{Planetary Collision Impact Scenario}

Our current interpretation of the available data is that Gaia-GIC-1 resembles a transiting planetary collision afterglow. There are only a handful of known stellar variables associated with anti-correlated near-IR and optical properties, a typical signature of newly generated circumstellar dust. In such instances, the freshly produced dust grains from the collisional debris absorb and scatter in the optical wavelengths, while re-radiating thermally in the infrared \citep{Wyatt_Jackson_Review}. The temporal evolution of photometric evolution in Gaia-GIC-1 suggests we may be witnessing the dynamic clearing and geometric evolution of impact-generated debris and the host star. Gaia-GIC-1 most clearly resembles the light curve properties of ASASSN-21qj \citep{Kenworthy_ASASSN21qj, Marshall_21qj}, with both systems exhibiting a striking transition from optical quiescent to irregular optical variability accompanied by the brightening in the near-IR. However, we note there are several distinguishing factors of this system. For instance, Gaia-GIC-1 shows three quasi-periodic optical dips before the onset of the irregular variability, which provide a strong constraint on the period of the transiting material to an equivalent semi-major axis of $\sim$1.1 AU. Such dusty occultation before the onset of the near-IR excess has been observed before, such as the case of HD 161991 \citep{Su_HD166191}, and other analog systems \citep{Moor_TYC_4209}. However, some alternative theories suggest that such complex variables could also be consistent with the breakup of exocometary bodies \citep{Marshall_21qj}. The large-amplitude and sudden increase in the near-IR flux points to a sudden large generation of a warm circumstellar disk in the inner region. 

Although the MIST age posteriors do not provide strong constraints on the stellar age, Gaia-GIC-1 displays several indicators that suggest a young age with the current available archival data. We examined the 2MASS color-color locus of Gaia-GIC-1 using its $(H-K_s)$ and $(H-J)$ colors. The target lies within $1\sigma$ of both the weak-line T-Tauri star (WTTS) and main-sequence loci, while appearing inconsistent with the classical T-Tauri star (CTTS) locus \citep{WWTS_disks, Meyer_97}. While the 2MASS colors alone cannot definitively distinguish a WTTS or main-sequence classification, disfavoring a CTTS classification suggests an age of $>10$Myr. Our tentative WTTS classification would be consistent with the ages of candidate host clusters FSR 1352/1347, which are generally above 10 Myr and are compatible with our EW$_{H\alpha}$ limits. Population-level statistics of protoplanetary disks suggest that the disk fraction among low-mass young stars drops to a few percent by $>$10 Myr in the 3.4-4.6 $\mu$m wavelength \citep{Ribas_2015}. Given the tentative constraints and clues on the stellar age, a planetary collision would be anticipated at this phase \citep{Wyatt_Jackson_Review}. At the present faint and highly irregular optical variability state, follow-up observations make it challenging to constrain any age diagnostic (i.e., joint constraint with SED and gyrochronological \citep{Angus_GyroSED} or Li I 6708 Å abundance \citep{EAGLES-Lithium}. Such constraints would likely only become feasible if the star returns to its quiescent brightness.

\begin{figure}
    \centering
    \includegraphics[width=1\linewidth]{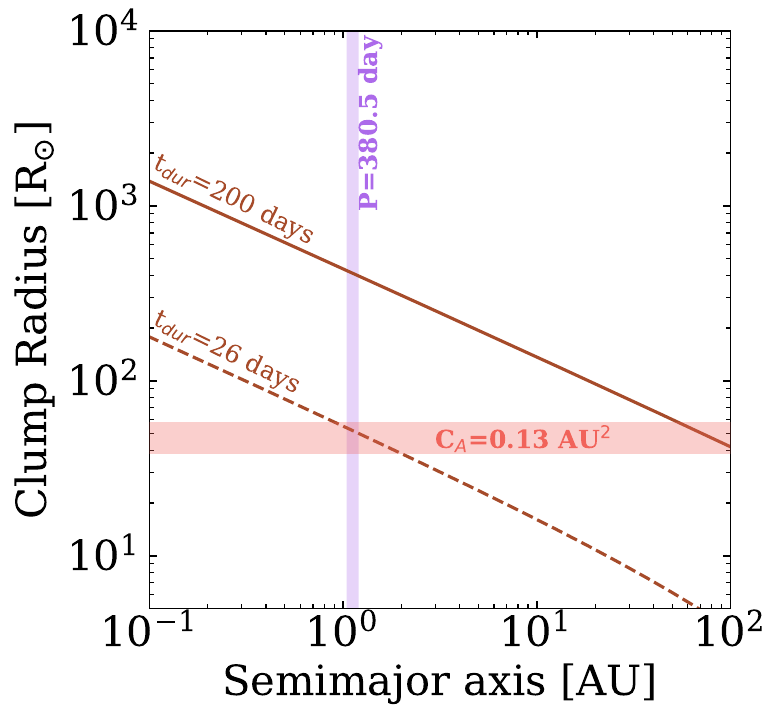}
    \caption{Radius of an occulting clump as a function of semi-major axis. Brown lines show the clump size required to match the observed transit durations (solid: 200 d; dashed: 26 d) for a circular, central transit around a 1.3M$_{\odot}$, 1.7 R$_{\odot}$ star. The horizontal pink line indicates the equivalent clump size implied by a total occulting cross-section of 0.13} AU$^{2}$. The vertical purple band shows the semi-major axis corresponding to an orbital period of 380.5 days, corresponding to a $\sim$1.1 AU orbit.
    \label{fig:clump_constraints}
\end{figure}

To quantify the debris mass produced by the proposed collision, we estimate the total dust mass based on the IR excess flux. From section \ref{sec:properties}, we adopt an average dust temperature of 900 Kelvin, and a total cross-sectional emitting area of 0.13 AU$^{2}$. Assuming the emission is due to small dust grains generated in a collisional cascade, with bulk density 3 g cm$^{-3}$ and a grain size distribution of 1 cm - 1 $\mu$m, we estimate the total emitting mass to be on the order of 5$\times 10^{21}$ kg. Our derived mass estimate is also consistent with the W$_{2}$ excess flux, which yields a similar cross-section of 0.08 AU$^{2}$. We caution that this must be a conservative minimum mass estimate since our distance error is on the order of $\sim$1 kpc, and the near-IR is probing the heated material within only a few AU \citep{Dullemond}. Our estimated dust mass is on the mass scale of a small icy moon, such as that of Enceladus \citep{RAPPAPORT2007175}, which likely indicates that the initial colliding bodies should be considerably larger, though this could be estimated more precisely through collisional simulations \citep{Kenworthy_ASASSN21qj}.

The observed transit properties of Gaia-GIC-1 appear to deviate from an edge-on circular orbit. In Figure \ref{fig:clump_constraints}, we show the anticipated clump radius assuming an edge-on circular orbit capable of producing the observed optical dips. The initial Gaia-G dips have depths of 25\% with characteristic durations of 200 days. At the constrained semi-major axis of 1.1 AU, this duration would imply a transiting dust cloud radius of $5\times 10^{2}$ $R_{\odot}$, requiring a substantially larger emitting cross-sectional area than the derived 0.13 AU$^{2}$. Conversely, if we adopt the measured cross-sectional area of 0.13 AU$^{2}$ at 1.1 AU, the implied transit duration would be only 26 days, which would be inconsistent with observations. We note that the degeneracy between semi-major axis, transit duration, and emitting area likely arises from an elongated dust structure, further supported by the highly asymmetric dimming profiles in the Gaia-G photometry. We constrain the transverse velocity of the occulting material by measuring flux slopes during ingress and egress, assuming the occultations are caused by a sharp dust edge \citep{2017RSOS....460652K}. We measure the steepest flux gradient with a transverse velocity of $3.0 \pm 0.4$ km s$^{-1}$, which is substantially lower than the anticipated $\sim$30 km s$^{-1}$ Keplerian velocity at 1.1 AU for a circular orbit. One scenario this could be consistent with is the debris being on a highly eccentric orbit, where the orbital velocity material is minimized near the apoapsis. Such orbital geometry could explain both the extended transit durations and the complex quasi-periodic behavior observed in Gaia-GIC-1. Alternatively, the extended transit durations and asymmetric dip profiles could arise from a gradual distribution of optical depth within the dust cloud itself, rather than being dominated by the physical velocity of the debris, which would be more consistent with the observed dimming profiles. Qualitatively, this scenario suggests an evolving, shearing debris field in which impact ejecta with high velocity dispersion will contribute to a range of semi-major axes a, as predicted by simulations \citep{Wyatt_Jackson_Review, Kenyon_2005}.

\subsection{Alternative Scenarios}

Tidally disrupted bodies, including gas giants \citep{Li2010_gas_giant, Oberst_16b}, planets \citep{Vanderburg_WD1145, Rappaport_K22b}, and exocomets \citep{Rappaport_exocomet}, can reach separations with their host stars that can facilitate either ongoing tidal stripping and Roche lobe overflow, or complete breakup near the Roche limit, by producing strongly asymmetric dusty transits (e.g., \cite{Hon_2025}). We explore whether the Gaia-GIC-1 light curve is consistent with tidal disruption of a planetesimal generating warm debris; however, the current available data suggest this scenario is unlikely. At the given constrained mass and radius of the primary star, the Roche limit for a comet-like body ($\rho\sim 0.5\,\mathrm{g\,cm}^{-3}$) is $\sim$0.01 AU, while an Earth-like planet ($\rho\sim 5.5\,\mathrm{g\,cm}^{-3}$) would have an even smaller Roche limit of $\sim$0.004AU. In the scenario of circularly orbiting dusty material resulting from the disruption of a comet-like body, if the material is near the Roche limit, this would be inconsistent with the initial periodic occultations measured to be at $\sim$1.1 AU semi-major axis. Similarly, assuming the occulting cross-sectional area of 0.13 AU$^{2}$ and stellar radius, the velocity at the Roche limit $\sim$352 km sec$^{-1}$ would imply a transit duration of $\sim$2 days, which is in stark contrast with the individual 200 day transits we find prior to the onset of the IR-excess nor the estimated 3 km sec$^{-1}$ minimum velocity estimated via the steepest flux gradient. For a highly eccentric orbit (e=0.9) at the Roche limit, a comet-like body would have periapsis and apoapsis velocities of $\sim$5,000 km sec$^{-1}$ and $\sim$85 km sec$^{-1}$, respectively. The resulting transiting timescales would be severely undersampled by the 21-day average Gaia cadence, making this scenario difficult to constrain. Thus, it remains plausible that if the material is near the Roche limit on highly elliptical orbits, and would require higher photometric monitoring to test more concretely. 

Alternative pathways to produce near-IR excess have also been previously proposed, including gravitational stirring of a dense planetesimal belt by an unseen planet or low-mass companion, and delayed orbital instabilities can trigger collisions, yielding episodic dust production \citep{2021ApJ...918...71R, 2020PSJ.....1...18C}. We did not find any archival data to suggest that Gaia-GIC-1 is a wide binary system that could drive such orbital instabilities. For example, we examined the Gaia RUWE score to be $\sim$1, and the nearby stellar source (Gaia DR3 5593847977064283520) is unlikely to be associated with Gaia-GIC-1, as they are 8.3 arcseconds apart and have a large physical separation.

Canonical T-Tauri stellar variables have also been found to have irregular variability in the optical and near-IR \citep{AMC_YSO_CAROT}. Though typically, these sources are known to have typical signatures or IR excess due to the presence of a circumstellar disk, signatures of their youth in an optical spectrum, such as signs of ongoing extreme accretion events through the presence of H$\alpha$ emission \citep{2025ApJ...988...77H}. Our spectroscopic H$\alpha$ EW upper limits would also suggest the unlikely scenario of dealing with a CTT star, though, within our detection limits, a more mature WTT star is plausible with typical EW$_{H_{\alpha}}$$<10$Å \citep{Barentsen2011}. Given our tentative optical spectrum, we rule out the scenario of a typical Class I/II YSO variability, since we did not find any such emission lines. Systematic searches for dipper variables among both young stellar objects \citep{Hillenbrand22} and main-sequence stars \citep{Tzan2025} have revealed diverse dimming phenomenologies. However, the combination of periodic and asymmetric optical dips, anti-correlated near-IR brightening, and lack of persistent IR excess in Gaia-GIC-1 distinguishes it from typical dipper populations, which generally lack significant IR variability and excess.

\section{Conclusion} \label{sec:conclusion}

We report the discovery of Gaia-GIC-1 (Gaia20ehk), an ongoing optical dimming and low-luminosity infrared stellar transient, which we hypothesize to be undergoing planetesimal collision afterglow seen by Gaia. Based on the constructed archival SED of the source before the irregular dimming events, we constrain the star to be a young F5-type star. Optical follow-up spectroscopy reveals a nearly featureless spectrum, which is likely due to the low SNR and high stellar variability obscured by dust, though we note that no prominent features of stellar accretion were found. Analysis of the Gaia-G optical light curve reveals a quasi-periodic optical dip with $\sim$25\% flux amplitude and timescale of 200 days with a period of $\sim$380.5 days, implying an orbital semi-major axis of 1.1 AU. Based on the available photometry, we derive a conservative dust mass of 4$\times 10^{20}$ kg, and dust temperature of 900 Kelvin, and suggest that the newly generated circumstellar material is likely from the collision of two planetesimals. Continued optical and near-IR photometry is strongly encouraged, as recent photometric optical-to-near-IR photometric detections show the stellar source to still exhibit strong variability.  

Continued infrared monitoring of Gaia-GIC-1 with the exceptional sensitivity and broad wavelength coverage of \textit{JWST} is essential for disentangling the origin of its mid‑IR excess and tracing the thermal evolution of any post‑impact debris. Multi‑epoch MIRI photometry and low‑resolution spectroscopy can pin down the temperature–luminosity decay curve of the warm dust, distinguishing a rapidly cooling impact afterglow from passively heated circumstellar material, and possibly resolve key solid‑state features such as the 9–12$\mu m$ silicate features that could be even more compelling evidence for a giant impact scenario (e.g., \cite{Su_2025_JWST}). Integrating time‑resolved \textit{JWST} observations with existing optical and NEOWISE light curves will allow construction of a self‑consistent model of the collision geometry, debris mass, and clearing timescales. Thanks to the Gaia Photometric Science Alerts, we have likely found the first planetesimal impact seen by Gaia and other space-based observatories due to its all-sky photometric coverage over the last decade. Subsequent discoveries can also be driven by the Vera C. Rubin Observatory's Legacy Survey of Space and Time (LSST) \citep{2019ApJ...873..111I}. 

\
\

\section{Acknowledgments}
The authors thank the anonymous referee for their helpful comments and suggestions that improved the quality of this work. AT acknowledges useful conversations with Eric Gaidos, Lynne Hillenbrand, and Nick Tusay for understanding the origins of the following system. We thank Chung-Uk Lee for obtaining follow-up photometry and Kishalay De for helping us with the unWISE image differencing analysis. The authors acknowledge support from the Institute for Data Intensive Research in Astrophysics \& Cosmology (DiRAC) in the Department of Astronomy at the University of Washington. The DiRAC Institute is supported through generous gifts from the Charles and Lisa Simonyi Fund for Arts and Sciences and the Washington Research Foundation. The authors would also like to acknowledge the generous support from Breakthrough Listen. The Breakthrough Prize Foundation funds the Breakthrough Initiatives, which manage Breakthrough Listen. This publication makes use of data products from the Wide-field Infrared Survey Explorer and NASA/IPAC Infrared Science Archive, which is a joint project of the University of California, Los Angeles, and the Jet Propulsion Laboratory/California Institute of Technology, funded by the National Aeronautics and Space Administration. This publication makes use of data products from the Spectro-Photometer for the History of the Universe, Epoch of Reionization and Ices Explorer (SPHEREx), which is a joint project of the Jet Propulsion Laboratory and the California Institute of Technology, and is funded by the National Aeronautics and Space Administration. This research has made use of publicly available data (https://kmtnet.kasi.re.kr/ulens/) from the KMTNet system operated by the Korea Astronomy and Space Science Institute (KASI) at three host sites of CTIO in Chile, SAAO in South Africa, and SSO in Australia. Data transfer from the host site to KASI was supported by the Korea Research Environment Open NETwork (KREONET).

\bibliography{sample631}{}

@ARTICLE{KMTNET,
       author = {{Kim}, Seung-Lee and {Lee}, Chung-Uk and {Park}, Byeong-Gon and {Kim}, Dong-Jin and {Cha}, Sang-Mok and {Lee}, Yongseok and {Han}, Cheongho and {Chun}, Moo-Young and {Yuk}, Insoo},
        title = "{KMTNET: A Network of 1.6 m Wide-Field Optical Telescopes Installed at Three Southern Observatories}",
      journal = {Journal of Korean Astronomical Society},
         year = 2016,
        month = feb,
       volume = {49},
       number = {1},
        pages = {37-44},
          doi = {10.5303/JKAS.2016.49.1.37},
       adsurl = {https://ui.adsabs.harvard.edu/abs/2016JKAS...49...37K}
}

@software{2015ascl.soft03010M,
       author = {{Morton}, Timothy D.},
        title = "{isochrones: Stellar model grid package}",
 howpublished = {Astrophysics Source Code Library, record ascl:1503.010},
         year = 2015,
        month = mar,
          eid = {ascl:1503.010},
       adsurl = {https://ui.adsabs.harvard.edu/abs/2015ascl.soft03010M}
}

@ARTICLE{SFD_dust,
       author = {{Schlegel}, David J. and {Finkbeiner}, Douglas P. and {Davis}, Marc},
        title = "{Maps of Dust Infrared Emission for Use in Estimation of Reddening and Cosmic Microwave Background Radiation Foregrounds}",
      journal = {\apj},
         year = 1998,
        month = jun,
       volume = {500},
       number = {2},
        pages = {525-553},
          doi = {10.1086/305772},
archivePrefix = {arXiv},
       eprint = {astro-ph/9710327},
 primaryClass = {astro-ph},
       adsurl = {https://ui.adsabs.harvard.edu/abs/1998ApJ...500..525S}
}

@ARTICLE{Tzan2025,
       author = {{Tzanidakis}, Anastasios and {Davenport}, James R.~A. and {Caplar}, Neven and {Bellm}, Eric C. and {Beebe}, Wilson and {Branton}, Doug and {Campos}, Sandro and {Connolly}, Andrew J. and {DeLucchi}, Melissa and {Malanchev}, Konstantin and {McGuire}, Sean},
        title = "{A Systematic Search for Main-sequence Dipper Stars Using the Zwicky Transient Facility}",
      journal = {\apj},
         year = 2025,
        month = sep,
       volume = {991},
       number = {1},
          eid = {118},
        pages = {118},
          doi = {10.3847/1538-4357/adf5bb},
archivePrefix = {arXiv},
       eprint = {2508.03964},
 primaryClass = {astro-ph.SR},
       adsurl = {https://ui.adsabs.harvard.edu/abs/2025ApJ...991..118T}
}

@ARTICLE{BailerJonesDR3,
       author = {{Bailer-Jones}, C.~A.~L. and {Rybizki}, J. and {Fouesneau}, M. and {Demleitner}, M. and {Andrae}, R.},
        title = "{Estimating Distances from Parallaxes. V. Geometric and Photogeometric Distances to 1.47 Billion Stars in Gaia Early Data Release 3}",
      journal = {\aj},
         year = 2021,
        month = mar,
       volume = {161},
       number = {3},
          eid = {147},
        pages = {147},
          doi = {10.3847/1538-3881/abd806},
archivePrefix = {arXiv},
       eprint = {2012.05220},
 primaryClass = {astro-ph.SR},
       adsurl = {https://ui.adsabs.harvard.edu/abs/2021AJ....161..147B}
}

@ARTICLE{2009MNRAS.398.1601F,
       author = {{Feroz}, F. and {Hobson}, M.~P. and {Bridges}, M.},
        title = "{MULTINEST: an efficient and robust Bayesian inference tool for cosmology and particle physics}",
      journal = {\mnras},
         year = 2009,
        month = oct,
       volume = {398},
       number = {4},
        pages = {1601-1614},
          doi = {10.1111/j.1365-2966.2009.14548.x},
archivePrefix = {arXiv},
       eprint = {0809.3437},
 primaryClass = {astro-ph},
       adsurl = {https://ui.adsabs.harvard.edu/abs/2009MNRAS.398.1601F}
}

@ARTICLE{decaps,
       author = {{Schlafly}, E.~F. and {Green}, G.~M. and {Lang}, D. and {Daylan}, T. and {Finkbeiner}, D.~P. and {Lee}, A. and {Meisner}, A.~M. and {Schlegel}, D. and {Valdes}, F.},
        title = "{The DECam Plane Survey: Optical Photometry of Two Billion Objects in the Southern Galactic Plane}",
      journal = {\apjs},
         year = 2018,
        month = feb,
       volume = {234},
       number = {2},
          eid = {39},
        pages = {39},
          doi = {10.3847/1538-4365/aaa3e2},
archivePrefix = {arXiv},
       eprint = {1710.01309},
 primaryClass = {astro-ph.GA},
       adsurl = {https://ui.adsabs.harvard.edu/abs/2018ApJS..234...39S}
}

@ARTICLE{skymap_dr4,
       author = {{Onken}, Christopher A. and {Wolf}, Christian and {Bessell}, Michael S. and {Chang}, Seo-Won and {Luvaul}, Lance C. and {Tonry}, John L. and {White}, Marc C. and {Da Costa}, Gary S.},
        title = "{SkyMapper Southern Survey: Data release 4}",
      journal = {\pasa},
         year = 2024,
        month = oct,
       volume = {41},
          eid = {e061},
        pages = {e061},
          doi = {10.1017/pasa.2024.53},
archivePrefix = {arXiv},
       eprint = {2402.02015},
 primaryClass = {astro-ph.CO},
       adsurl = {https://ui.adsabs.harvard.edu/abs/2024PASA...41...61O}
}

@ARTICLE{Cody18,
       author = {{Cody}, Ann Marie and {Hillenbrand}, Lynne A.},
        title = "{The Many-faceted Light Curves of Young Disk-bearing Stars in Upper Sco -- Oph Observed by K2 Campaign 2}",
      journal = {\aj},
         year = 2018,
        month = aug,
       volume = {156},
       number = {2},
          eid = {71},
        pages = {71},
          doi = {10.3847/1538-3881/aacead},
archivePrefix = {arXiv},
       eprint = {1802.06409},
 primaryClass = {astro-ph.SR},
       adsurl = {https://ui.adsabs.harvard.edu/abs/2018AJ....156...71C}
}

@ARTICLE{Cody14,
       author = {{Cody}, Ann Marie and {Stauffer}, John and {Baglin}, Annie and {Micela}, Giuseppina and {Rebull}, Luisa M. and {Flaccomio}, Ettore and {Morales-Calder{\'o}n}, Mar{\'\i}a and {Aigrain}, Suzanne and {Bouvier}, J{\`e}r{\^o}me and {Hillenbrand}, Lynne A. and {Gutermuth}, Robert and {Song}, Inseok and {Turner}, Neal and {Alencar}, Silvia H.~P. and {Zwintz}, Konstanze and {Plavchan}, Peter and {Carpenter}, John and {Findeisen}, Krzysztof and {Carey}, Sean and {Terebey}, Susan and {Hartmann}, Lee and {Calvet}, Nuria and {Teixeira}, Paula and {Vrba}, Frederick J. and {Wolk}, Scott and {Covey}, Kevin and {Poppenhaeger}, Katja and {G{\"u}nther}, Hans Moritz and {Forbrich}, Jan and {Whitney}, Barbara and {Affer}, Laura and {Herbst}, William and {Hora}, Joseph and {Barrado}, David and {Holtzman}, Jon and {Marchis}, Franck and {Wood}, Kenneth and {Medeiros Guimar{\~a}es}, Marcelo and {Lillo Box}, Jorge and {Gillen}, Ed and {McQuillan}, Amy and {Espaillat}, Catherine and {Allen}, Lori and {D'Alessio}, Paola and {Favata}, Fabio},
        title = "{CSI 2264: Simultaneous Optical and Infrared Light Curves of Young Disk-bearing Stars in NGC 2264 with CoRoT and Spitzer{\textemdash}Evidence for Multiple Origins of Variability}",
      journal = {\aj},
         year = 2014,
        month = apr,
       volume = {147},
       number = {4},
          eid = {82},
        pages = {82},
          doi = {10.1088/0004-6256/147/4/82},
archivePrefix = {arXiv},
       eprint = {1401.6582},
 primaryClass = {astro-ph.SR},
       adsurl = {https://ui.adsabs.harvard.edu/abs/2014AJ....147...82C}
}

@ARTICLE{Rappaport_exocomet,
       author = {{Rappaport}, S. and {Vanderburg}, A. and {Jacobs}, T. and {LaCourse}, D. and {Jenkins}, J. and {Kraus}, A. and {Rizzuto}, A. and {Latham}, D.~W. and {Bieryla}, A. and {Lazarevic}, M. and {Schmitt}, A.},
        title = "{Likely transiting exocomets detected by Kepler}",
      journal = {\mnras},
         year = 2018,
        month = feb,
       volume = {474},
       number = {2},
        pages = {1453-1468},
          doi = {10.1093/mnras/stx2735},
archivePrefix = {arXiv},
       eprint = {1708.06069},
 primaryClass = {astro-ph.EP},
       adsurl = {https://ui.adsabs.harvard.edu/abs/2018MNRAS.474.1453R}
}

@ARTICLE{Rappaport_K22b,
       author = {{Rappaport}, S. and {Levine}, A. and {Chiang}, E. and {El Mellah}, I. and {Jenkins}, J. and {Kalomeni}, B. and {Kite}, E.~S. and {Kotson}, M. and {Nelson}, L. and {Rousseau-Nepton}, L. and {Tran}, K.},
        title = "{Possible Disintegrating Short-period Super-Mercury Orbiting KIC 12557548}",
      journal = {\apj},
         year = 2012,
        month = jun,
       volume = {752},
       number = {1},
          eid = {1},
        pages = {1},
          doi = {10.1088/0004-637X/752/1/1},
archivePrefix = {arXiv},
       eprint = {1201.2662},
 primaryClass = {astro-ph.EP},
       adsurl = {https://ui.adsabs.harvard.edu/abs/2012ApJ...752....1R}
}

@ARTICLE{Kenworthy_ASASSN21qj,
       author = {{Kenworthy}, Matthew and {Lock}, Simon and {Kennedy}, Grant and {van Capelleveen}, Richelle and {Mamajek}, Eric and {Carone}, Ludmila and {Hambsch}, Franz-Josef and {Masiero}, Joseph and {Mainzer}, Amy and {Kirkpatrick}, J. Davy and {Gomez}, Edward and {Leinhardt}, Zo{\"e} and {Dou}, Jingyao and {Tanna}, Pavan and {Sainio}, Arttu and {Barker}, Hamish and {Charbonnel}, St{\'e}phane and {Garde}, Olivier and {Le D{\^u}}, Pascal and {Mulato}, Lionel and {Petit}, Thomas and {Rizzo Smith}, Michael},
        title = "{A planetary collision afterglow and transit of the resultant debris cloud}",
      journal = {\nat},
         year = 2023,
        month = oct,
       volume = {622},
       number = {7982},
        pages = {251-254},
 primaryClass = {astro-ph.EP},
       adsurl = {https://ui.adsabs.harvard.edu/abs/2023Natur.622..251K}
}

@ARTICLE{2019ApJ...873..111I,
       author = {{Ivezi{\'c}}, {\v{Z}}eljko and {Kahn}, Steven M. and {Tyson}, J. Anthony and {Abel}, Bob and {Acosta}, Emily and {Allsman}, Robyn and {Alonso}, David and {AlSayyad}, Yusra and {Anderson}, Scott F. and {Andrew}, John and {Angel}, James Roger P. and {Angeli}, George Z. and {Ansari}, Reza and {Antilogus}, Pierre and {Araujo}, Constanza and {Armstrong}, Robert and {Arndt}, Kirk T. and {Astier}, Pierre and {Aubourg}, {\'E}ric and {Auza}, Nicole and {Axelrod}, Tim S. and {Bard}, Deborah J. and {Barr}, Jeff D. and {Barrau}, Aurelian and {Bartlett}, James G. and {Bauer}, Amanda E. and {Bauman}, Brian J. and {Baumont}, Sylvain and {Bechtol}, Ellen and {Bechtol}, Keith and {Becker}, Andrew C. and {Becla}, Jacek and {Beldica}, Cristina and {Bellavia}, Steve and {Bianco}, Federica B. and {Biswas}, Rahul and {Blanc}, Guillaume and {Blazek}, Jonathan and {Blandford}, Roger D. and {Bloom}, Josh S. and {Bogart}, Joanne and {Bond}, Tim W. and {Booth}, Michael T. and {Borgland}, Anders W. and {Borne}, Kirk and {Bosch}, James F. and {Boutigny}, Dominique and {Brackett}, Craig A. and {Bradshaw}, Andrew and {Brandt}, William Nielsen and {Brown}, Michael E. and {Bullock}, James S. and {Burchat}, Patricia and {Burke}, David L. and {Cagnoli}, Gianpietro and {Calabrese}, Daniel and {Callahan}, Shawn and {Callen}, Alice L. and {Carlin}, Jeffrey L. and {Carlson}, Erin L. and {Chandrasekharan}, Srinivasan and {Charles-Emerson}, Glenaver and {Chesley}, Steve and {Cheu}, Elliott C. and {Chiang}, Hsin-Fang and {Chiang}, James and {Chirino}, Carol and {Chow}, Derek and {Ciardi}, David R. and {Claver}, Charles F. and {Cohen-Tanugi}, Johann and {Cockrum}, Joseph J. and {Coles}, Rebecca and {Connolly}, Andrew J. and {Cook}, Kem H. and {Cooray}, Asantha and {Covey}, Kevin R. and {Cribbs}, Chris and {Cui}, Wei and {Cutri}, Roc and {Daly}, Philip N. and {Daniel}, Scott F. and {Daruich}, Felipe and {Daubard}, Guillaume and {Daues}, Greg and {Dawson}, William and {Delgado}, Francisco and {Dellapenna}, Alfred and {de Peyster}, Robert and {de Val-Borro}, Miguel and {Digel}, Seth W. and {Doherty}, Peter and {Dubois}, Richard and {Dubois-Felsmann}, Gregory P. and {Durech}, Josef and {Economou}, Frossie and {Eifler}, Tim and {Eracleous}, Michael and {Emmons}, Benjamin L. and {Fausti Neto}, Angelo and {Ferguson}, Henry and {Figueroa}, Enrique and {Fisher-Levine}, Merlin and {Focke}, Warren and {Foss}, Michael D. and {Frank}, James and {Freemon}, Michael D. and {Gangler}, Emmanuel and {Gawiser}, Eric and {Geary}, John C. and {Gee}, Perry and {Geha}, Marla and {Gessner}, Charles J.~B. and {Gibson}, Robert R. and {Gilmore}, D. Kirk and {Glanzman}, Thomas and {Glick}, William and {Goldina}, Tatiana and {Goldstein}, Daniel A. and {Goodenow}, Iain and {Graham}, Melissa L. and {Gressler}, William J. and {Gris}, Philippe and {Guy}, Leanne P. and {Guyonnet}, Augustin and {Haller}, Gunther and {Harris}, Ron and {Hascall}, Patrick A. and {Haupt}, Justine and {Hernandez}, Fabio and {Herrmann}, Sven and {Hileman}, Edward and {Hoblitt}, Joshua and {Hodgson}, John A. and {Hogan}, Craig and {Howard}, James D. and {Huang}, Dajun and {Huffer}, Michael E. and {Ingraham}, Patrick and {Innes}, Walter R. and {Jacoby}, Suzanne H. and {Jain}, Bhuvnesh and {Jammes}, Fabrice and {Jee}, M. James and {Jenness}, Tim and {Jernigan}, Garrett and {Jevremovi{\'c}}, Darko and {Johns}, Kenneth and {Johnson}, Anthony S. and {Johnson}, Margaret W.~G. and {Jones}, R. Lynne and {Juramy-Gilles}, Claire and {Juri{\'c}}, Mario and {Kalirai}, Jason S. and {Kallivayalil}, Nitya J. and {Kalmbach}, Bryce and {Kantor}, Jeffrey P. and {Karst}, Pierre and {Kasliwal}, Mansi M. and {Kelly}, Heather and {Kessler}, Richard and {Kinnison}, Veronica and {Kirkby}, David and {Knox}, Lloyd and {Kotov}, Ivan V. and {Krabbendam}, Victor L. and {Krughoff}, K. Simon and {Kub{\'a}nek}, Petr and {Kuczewski}, John and {Kulkarni}, Shri and {Ku}, John and {Kurita}, Nadine R. and {Lage}, Craig S. and {Lambert}, Ron and {Lange}, Travis and {Langton}, J. Brian and {Le Guillou}, Laurent and {Levine}, Deborah and {Liang}, Ming and {Lim}, Kian-Tat and {Lintott}, Chris J. and {Long}, Kevin E. and {Lopez}, Margaux and {Lotz}, Paul J. and {Lupton}, Robert H. and {Lust}, Nate B. and {MacArthur}, Lauren A. and {Mahabal}, Ashish and {Mandelbaum}, Rachel and {Markiewicz}, Thomas W. and {Marsh}, Darren S. and {Marshall}, Philip J. and {Marshall}, Stuart and {May}, Morgan and {McKercher}, Robert and {McQueen}, Michelle and {Meyers}, Joshua and {Migliore}, Myriam and {Miller}, Michelle and {Mills}, David J. and {Miraval}, Connor and {Moeyens}, Joachim and {Moolekamp}, Fred E. and {Monet}, David G. and {Moniez}, Marc and {Monkewitz}, Serge and {Montgomery}, Christopher and {Morrison}, Christopher B. and {Mueller}, Fritz and {Muller}, Gary P. and {Mu{\~n}oz Arancibia}, Freddy and {Neill}, Douglas R. and {Newbry}, Scott P. and {Nief}, Jean-Yves and {Nomerotski}, Andrei and {Nordby}, Martin and {O'Connor}, Paul and {Oliver}, John and {Olivier}, Scot S. and {Olsen}, Knut and {O'Mullane}, William and {Ortiz}, Sandra and {Osier}, Shawn and {Owen}, Russell E. and {Pain}, Reynald and {Palecek}, Paul E. and {Parejko}, John K. and {Parsons}, James B. and {Pease}, Nathan M. and {Peterson}, J. Matt and {Peterson}, John R. and {Petravick}, Donald L. and {Libby Petrick}, M.~E. and {Petry}, Cathy E. and {Pierfederici}, Francesco and {Pietrowicz}, Stephen and {Pike}, Rob and {Pinto}, Philip A. and {Plante}, Raymond and {Plate}, Stephen and {Plutchak}, Joel P. and {Price}, Paul A. and {Prouza}, Michael and {Radeka}, Veljko and {Rajagopal}, Jayadev and {Rasmussen}, Andrew P. and {Regnault}, Nicolas and {Reil}, Kevin A. and {Reiss}, David J. and {Reuter}, Michael A. and {Ridgway}, Stephen T. and {Riot}, Vincent J. and {Ritz}, Steve and {Robinson}, Sean and {Roby}, William and {Roodman}, Aaron and {Rosing}, Wayne and {Roucelle}, Cecille and {Rumore}, Matthew R. and {Russo}, Stefano and {Saha}, Abhijit and {Sassolas}, Benoit and {Schalk}, Terry L. and {Schellart}, Pim and {Schindler}, Rafe H. and {Schmidt}, Samuel and {Schneider}, Donald P. and {Schneider}, Michael D. and {Schoening}, William and {Schumacher}, German and {Schwamb}, Megan E. and {Sebag}, Jacques and {Selvy}, Brian and {Sembroski}, Glenn H. and {Seppala}, Lynn G. and {Serio}, Andrew and {Serrano}, Eduardo and {Shaw}, Richard A. and {Shipsey}, Ian and {Sick}, Jonathan and {Silvestri}, Nicole and {Slater}, Colin T. and {Smith}, J. Allyn and {Smith}, R. Chris and {Sobhani}, Shahram and {Soldahl}, Christine and {Storrie-Lombardi}, Lisa and {Stover}, Edward and {Strauss}, Michael A. and {Street}, Rachel A. and {Stubbs}, Christopher W. and {Sullivan}, Ian S. and {Sweeney}, Donald and {Swinbank}, John D. and {Szalay}, Alexander and {Takacs}, Peter and {Tether}, Stephen A. and {Thaler}, Jon J. and {Thayer}, John Gregg and {Thomas}, Sandrine and {Thornton}, Adam J. and {Thukral}, Vaikunth and {Tice}, Jeffrey and {Trilling}, David E. and {Turri}, Max and {Van Berg}, Richard and {Vanden Berk}, Daniel and {Vetter}, Kurt and {Virieux}, Francoise and {Vucina}, Tomislav and {Wahl}, William and {Walkowicz}, Lucianne and {Walsh}, Brian and {Walter}, Christopher W. and {Wang}, Daniel L. and {Wang}, Shin-Yawn and {Warner}, Michael and {Wiecha}, Oliver and {Willman}, Beth and {Winters}, Scott E. and {Wittman}, David and {Wolff}, Sidney C. and {Wood-Vasey}, W. Michael and {Wu}, Xiuqin and {Xin}, Bo and {Yoachim}, Peter and {Zhan}, Hu},
        title = "{LSST: From Science Drivers to Reference Design and Anticipated Data Products}",
      journal = {\apj},
         year = 2019,
        month = mar,
       volume = {873},
       number = {2},
          eid = {111},
        pages = {111},
          doi = {10.3847/1538-4357/ab042c},
archivePrefix = {arXiv},
       eprint = {0805.2366},
 primaryClass = {astro-ph},
       adsurl = {https://ui.adsabs.harvard.edu/abs/2019ApJ...873..111I}
}

@ARTICLE{2012MNRAS.425..657J,
       author = {{Jackson}, Alan P. and {Wyatt}, Mark C.},
        title = "{Debris from terrestrial planet formation: the Moon-forming collision}",
      journal = {\mnras},
         year = 2012,
        month = sep,
       volume = {425},
       number = {1},
        pages = {657-679},
          doi = {10.1111/j.1365-2966.2012.21546.x},
archivePrefix = {arXiv},
       eprint = {1206.4190},
 primaryClass = {astro-ph.EP},
       adsurl = {https://ui.adsabs.harvard.edu/abs/2012MNRAS.425..657J}
}

@ARTICLE{2023ApJ...955...69T,
       author = {{Tzanidakis}, Anastasios and {Davenport}, James R.~A. and {Bellm}, Eric C. and {Wang}, Yuankun},
        title = "{Gaia17bpp: A Giant Star with the Deepest and Longest Known Dimming Event}",
      journal = {\apj},
         year = 2023,
        month = sep,
       volume = {955},
       number = {1},
          eid = {69},
        pages = {69},
          doi = {10.3847/1538-4357/aceda7},
archivePrefix = {arXiv},
       eprint = {2306.12409},
 primaryClass = {astro-ph.SR},
       adsurl = {https://ui.adsabs.harvard.edu/abs/2023ApJ...955...69T}
}

@ARTICLE{2mass,
       author = {{Skrutskie}, M.~F. and {Cutri}, R.~M. and {Stiening}, R. and {Weinberg}, M.~D. and {Schneider}, S. and {Carpenter}, J.~M. and {Beichman}, C. and {Capps}, R. and {Chester}, T. and {Elias}, J. and {Huchra}, J. and {Liebert}, J. and {Lonsdale}, C. and {Monet}, D.~G. and {Price}, S. and {Seitzer}, P. and {Jarrett}, T. and {Kirkpatrick}, J.~D. and {Gizis}, J.~E. and {Howard}, E. and {Evans}, T. and {Fowler}, J. and {Fullmer}, L. and {Hurt}, R. and {Light}, R. and {Kopan}, E.~L. and {Marsh}, K.~A. and {McCallon}, H.~L. and {Tam}, R. and {Van Dyk}, S. and {Wheelock}, S.},
        title = "{The Two Micron All Sky Survey (2MASS)}",
      journal = {\aj},
         year = 2006,
        month = feb,
       volume = {131},
       number = {2},
        pages = {1163-1183},
          doi = {10.1086/498708},
       adsurl = {https://ui.adsabs.harvard.edu/abs/2006AJ....131.1163S}
}

@ARTICLE{Su_HD166191,
       author = {{Su}, Kate Y.~L. and {Kennedy}, Grant M. and {Schlawin}, Everett and {Jackson}, Alan P. and {Rieke}, G.~H.},
        title = "{A Star-sized Impact-produced Dust Clump in the Terrestrial Zone of the HD 166191 System}",
      journal = {\apj},
         year = 2022,
        month = mar,
       volume = {927},
       number = {2},
          eid = {135},
        pages = {135},
          doi = {10.3847/1538-4357/ac4bbb},
archivePrefix = {arXiv},
       eprint = {2203.02366},
 primaryClass = {astro-ph.EP},
       adsurl = {https://ui.adsabs.harvard.edu/abs/2022ApJ...927..135S}
}

@ARTICLE{Quintana_GI_freq,
       author = {{Quintana}, Elisa V. and {Barclay}, Thomas and {Borucki}, William J. and {Rowe}, Jason F. and {Chambers}, John E.},
        title = "{The Frequency of Giant Impacts on Earth-like Worlds}",
      journal = {\apj},
         year = 2016,
        month = apr,
       volume = {821},
       number = {2},
          eid = {126},
        pages = {126},
          doi = {10.3847/0004-637X/821/2/126},
archivePrefix = {arXiv},
       eprint = {1511.03663},
 primaryClass = {astro-ph.EP},
       adsurl = {https://ui.adsabs.harvard.edu/abs/2016ApJ...821..126Q}
}

@ARTICLE{LammersGI_ML,
       author = {{Lammers}, Caleb and {Cranmer}, Miles and {Hadden}, Sam and {Ho}, Shirley and {Murray}, Norman and {Tamayo}, Daniel},
        title = "{Accelerating Giant-impact Simulations with Machine Learning}",
      journal = {\apj},
         year = 2024,
        month = nov,
       volume = {975},
       number = {2},
          eid = {228},
        pages = {228},
          doi = {10.3847/1538-4357/ad7fe5},
archivePrefix = {arXiv},
       eprint = {2408.08873},
 primaryClass = {astro-ph.EP},
       adsurl = {https://ui.adsabs.harvard.edu/abs/2024ApJ...975..228L}
}

@ARTICLE{Steele_2016,
       author = {{Steele}, Amy and {Hughes}, A. Meredith and {Carpenter}, John and {Ricarte}, Angelo and {Andrews}, Sean M. and {Wilner}, David J. and {Chiang}, Eugene},
        title = "{Resolved Millimeter-wavelength Observations of Debris Disks around Solar-type Stars}",
      journal = {\apj},
         year = 2016,
        month = jan,
       volume = {816},
       number = {1},
          eid = {27},
        pages = {27},
          doi = {10.3847/0004-637X/816/1/27},
archivePrefix = {arXiv},
       eprint = {1510.08890},
 primaryClass = {astro-ph.EP},
       adsurl = {https://ui.adsabs.harvard.edu/abs/2016ApJ...816...27S}
}

@ARTICLE{unTimely,
       author = {{Meisner}, Aaron M. and {Caselden}, Dan and {Schlafly}, Edward F. and {Kiwy}, Frank},
        title = "{unTimely: a Full-sky, Time-domain unWISE Catalog}",
      journal = {\aj},
         year = 2023,
        month = feb,
       volume = {165},
       number = {2},
          eid = {36},
        pages = {36},
          doi = {10.3847/1538-3881/aca2ab},
archivePrefix = {arXiv},
       eprint = {2209.14327},
 primaryClass = {astro-ph.IM},
       adsurl = {https://ui.adsabs.harvard.edu/abs/2023AJ....165...36M}
}

@ARTICLE{GaiaAlerts_Hodgkin,
       author = {{Hodgkin}, S.~T. and {Harrison}, D.~L. and {Breedt}, E. and {Wevers}, T. and {Rixon}, G. and {Delgado}, A. and {Yoldas}, A. and {Kostrzewa-Rutkowska}, Z. and {Wyrzykowski}, {\L}. and {van Leeuwen}, M. and {Blagorodnova}, N. and {Campbell}, H. and {Eappachen}, D. and {Fraser}, M. and {Ihanec}, N. and {Koposov}, S.~E. and {Kruszy{\'n}ska}, K. and {Marton}, G. and {Rybicki}, K.~A. and {Brown}, A.~G.~A. and {Burgess}, P.~W. and {Busso}, G. and {Cowell}, S. and {De Angeli}, F. and {Diener}, C. and {Evans}, D.~W. and {Gilmore}, G. and {Holland}, G. and {Jonker}, P.~G. and {van Leeuwen}, F. and {Mignard}, F. and {Osborne}, P.~J. and {Portell}, J. and {Prusti}, T. and {Richards}, P.~J. and {Riello}, M. and {Seabroke}, G.~M. and {Walton}, N.~A. and {{\'A}brah{\'a}m}, P. and {Altavilla}, G. and {Baker}, S.~G. and {Bastian}, U. and {O'Brien}, P. and {de Bruijne}, J. and {Butterley}, T. and {Carrasco}, J.~M. and {Casta{\~n}eda}, J. and {Clark}, J.~S. and {Clementini}, G. and {Copperwheat}, C.~M. and {Cropper}, M. and {Damljanovic}, G. and {Davidson}, M. and {Davis}, C.~J. and {Dennefeld}, M. and {Dhillon}, V.~S. and {Dolding}, C. and {Dominik}, M. and {Esquej}, P. and {Eyer}, L. and {Fabricius}, C. and {Fridman}, M. and {Froebrich}, D. and {Garralda}, N. and {Gomboc}, A. and {Gonz{\'a}lez-Vidal}, J.~J. and {Guerra}, R. and {Hambly}, N.~C. and {Hardy}, L.~K. and {Holl}, B. and {Hourihane}, A. and {Japelj}, J. and {Kann}, D.~A. and {Kiss}, C. and {Knigge}, C. and {Kolb}, U. and {Komossa}, S. and {K{\'o}sp{\'a}l}, {\'A}. and {Kov{\'a}cs}, G. and {Kun}, M. and {Leto}, G. and {Lewis}, F. and {Littlefair}, S.~P. and {Mahabal}, A.~A. and {Mundell}, C.~G. and {Nagy}, Z. and {Padeletti}, D. and {Palaversa}, L. and {Pigulski}, A. and {Pretorius}, M.~L. and {van Reeven}, W. and {Ribeiro}, V.~A.~R.~M. and {Roelens}, M. and {Rowell}, N. and {Schartel}, N. and {Scholz}, A. and {Schwope}, A. and {Sip{\H{o}}cz}, B.~M. and {Smartt}, S.~J. and {Smith}, M.~D. and {Serraller}, I. and {Steeghs}, D. and {Sullivan}, M. and {Szabados}, L. and {Szegedi-Elek}, E. and {Tisserand}, P. and {Tomasella}, L. and {van Velzen}, S. and {Whitelock}, P.~A. and {Wilson}, R.~W. and {Young}, D.~R.},
        title = "{Gaia Early Data Release 3. Gaia photometric science alerts}",
      journal = {\aap},
         year = 2021,
        month = aug,
       volume = {652},
          eid = {A76},
        pages = {A76},
          doi = {10.1051/0004-6361/202140735},
archivePrefix = {arXiv},
       eprint = {2106.01394},
 primaryClass = {astro-ph.IM},
       adsurl = {https://ui.adsabs.harvard.edu/abs/2021A&A...652A..76H}
}

@ARTICLE{atel,
       author = {{Hodgkin}, S.~T. and {Breedt}, E. and {Delgado}, A. and {Harrison}, D.~L. and {Leeuwen}, M.~V. and {Rixon}, G. and {Wevers}, T. and {Yoldas}, A. and {Ihanec}, N. and {Kruszy{\'n}ska}, K. and {Rybicki}, K.~A. and {Wyrzykowski}, {\L}. and {Kostrzewa-Rutkowska}, Z. and {Eappachen}, D. and {Marton}, G.},
        title = "{GaiaAlerts Transient Discovery Report for 2020-09-14}",
      journal = {Transient Name Server Discovery Report},
         year = 2020,
        month = sep,
       volume = {2020-2792},
        pages = {1},
       adsurl = {https://ui.adsabs.harvard.edu/abs/2020TNSTR2792....1H}
}

@ARTICLE{2023A&A...674A...1G,
       author = {{Gaia Collaboration} and {Vallenari}, A. and {Brown}, A.~G.~A. and {Prusti}, T. and {de Bruijne}, J.~H.~J. and {Arenou}, F. and {Babusiaux}, C. and {Biermann}, M. and {Creevey}, O.~L. and {Ducourant}, C. and {Evans}, D.~W. and {Eyer}, L. and {Guerra}, R. and {Hutton}, A. and {Jordi}, C. and {Klioner}, S.~A. and {Lammers}, U.~L. and {Lindegren}, L. and {Luri}, X. and {Mignard}, F. and {Panem}, C. and {Pourbaix}, D. and {Randich}, S. and {Sartoretti}, P. and {Soubiran}, C. and {Tanga}, P. and {Walton}, N.~A. and {Bailer-Jones}, C.~A.~L. and {Bastian}, U. and {Drimmel}, R. and {Jansen}, F. and {Katz}, D. and {Lattanzi}, M.~G. and {van Leeuwen}, F. and {Bakker}, J. and {Cacciari}, C. and {Casta{\~n}eda}, J. and {De Angeli}, F. and {Fabricius}, C. and {Fouesneau}, M. and {Fr{\'e}mat}, Y. and {Galluccio}, L. and {Guerrier}, A. and {Heiter}, U. and {Masana}, E. and {Messineo}, R. and {Mowlavi}, N. and {Nicolas}, C. and {Nienartowicz}, K. and {Pailler}, F. and {Panuzzo}, P. and {Riclet}, F. and {Roux}, W. and {Seabroke}, G.~M. and {Sordo}, R. and {Th{\'e}venin}, F. and {Gracia-Abril}, G. and {Portell}, J. and {Teyssier}, D. and {Altmann}, M. and {Andrae}, R. and {Audard}, M. and {Bellas-Velidis}, I. and {Benson}, K. and {Berthier}, J. and {Blomme}, R. and {Burgess}, P.~W. and {Busonero}, D. and {Busso}, G. and {C{\'a}novas}, H. and {Carry}, B. and {Cellino}, A. and {Cheek}, N. and {Clementini}, G. and {Damerdji}, Y. and {Davidson}, M. and {de Teodoro}, P. and {Nu{\~n}ez Campos}, M. and {Delchambre}, L. and {Dell'Oro}, A. and {Esquej}, P. and {Fern{\'a}ndez-Hern{\'a}ndez}, J. and {Fraile}, E. and {Garabato}, D. and {Garc{\'\i}a-Lario}, P. and {Gosset}, E. and {Haigron}, R. and {Halbwachs}, J. -L. and {Hambly}, N.~C. and {Harrison}, D.~L. and {Hern{\'a}ndez}, J. and {Hestroffer}, D. and {Hodgkin}, S.~T. and {Holl}, B. and {Jan{\ss}en}, K. and {Jevardat de Fombelle}, G. and {Jordan}, S. and {Krone-Martins}, A. and {Lanzafame}, A.~C. and {L{\"o}ffler}, W. and {Marchal}, O. and {Marrese}, P.~M. and {Moitinho}, A. and {Muinonen}, K. and {Osborne}, P. and {Pancino}, E. and {Pauwels}, T. and {Recio-Blanco}, A. and {Reyl{\'e}}, C. and {Riello}, M. and {Rimoldini}, L. and {Roegiers}, T. and {Rybizki}, J. and {Sarro}, L.~M. and {Siopis}, C. and {Smith}, M. and {Sozzetti}, A. and {Utrilla}, E. and {van Leeuwen}, M. and {Abbas}, U. and {{\'A}brah{\'a}m}, P. and {Abreu Aramburu}, A. and {Aerts}, C. and {Aguado}, J.~J. and {Ajaj}, M. and {Aldea-Montero}, F. and {Altavilla}, G. and {{\'A}lvarez}, M.~A. and {Alves}, J. and {Anders}, F. and {Anderson}, R.~I. and {Anglada Varela}, E. and {Antoja}, T. and {Baines}, D. and {Baker}, S.~G. and {Balaguer-N{\'u}{\~n}ez}, L. and {Balbinot}, E. and {Balog}, Z. and {Barache}, C. and {Barbato}, D. and {Barros}, M. and {Barstow}, M.~A. and {Bartolom{\'e}}, S. and {Bassilana}, J. -L. and {Bauchet}, N. and {Becciani}, U. and {Bellazzini}, M. and {Berihuete}, A. and {Bernet}, M. and {Bertone}, S. and {Bianchi}, L. and {Binnenfeld}, A. and {Blanco-Cuaresma}, S. and {Blazere}, A. and {Boch}, T. and {Bombrun}, A. and {Bossini}, D. and {Bouquillon}, S. and {Bragaglia}, A. and {Bramante}, L. and {Breedt}, E. and {Bressan}, A. and {Brouillet}, N. and {Brugaletta}, E. and {Bucciarelli}, B. and {Burlacu}, A. and {Butkevich}, A.~G. and {Buzzi}, R. and {Caffau}, E. and {Cancelliere}, R. and {Cantat-Gaudin}, T. and {Carballo}, R. and {Carlucci}, T. and {Carnerero}, M.~I. and {Carrasco}, J.~M. and {Casamiquela}, L. and {Castellani}, M. and {Castro-Ginard}, A. and {Chaoul}, L. and {Charlot}, P. and {Chemin}, L. and {Chiaramida}, V. and {Chiavassa}, A. and {Chornay}, N. and {Comoretto}, G. and {Contursi}, G. and {Cooper}, W.~J. and {Cornez}, T. and {Cowell}, S. and {Crifo}, F. and {Cropper}, M. and {Crosta}, M. and {Crowley}, C. and {Dafonte}, C. and {Dapergolas}, A. and {David}, M. and {David}, P. and {de Laverny}, P. and {De Luise}, F. and {De March}, R.},
        title = "{Gaia Data Release 3. Summary of the content and survey properties}",
      journal = {\aap},
         year = 2023,
        month = jun,
       volume = {674},
          eid = {A1},
        pages = {A1},
          doi = {10.1051/0004-6361/202243940},
archivePrefix = {arXiv},
       eprint = {2208.00211},
 primaryClass = {astro-ph.GA},
       adsurl = {https://ui.adsabs.harvard.edu/abs/2023A&A...674A...1G}
}

@ARTICLE{pypiet,
       author = {{Prochaska}, J. and {Hennawi}, Joseph and {Westfall}, Kyle and {Cooke}, Ryan and {Wang}, Feige and {Hsyu}, Tiffany and {Davies}, Frederick and {Farina}, Emanuele and {Pelliccia}, Debora},
        title = "{PypeIt: The Python Spectroscopic Data Reduction Pipeline}",
      journal = {The Journal of Open Source Software},
         year = 2020,
        month = dec,
       volume = {5},
       number = {56},
          eid = {2308},
        pages = {2308},
          doi = {10.21105/joss.02308},
archivePrefix = {arXiv},
       eprint = {2005.06505},
 primaryClass = {astro-ph.IM},
       adsurl = {https://ui.adsabs.harvard.edu/abs/2020JOSS....5.2308P}
}

@ARTICLE{Jackson2014_planet_embroys,
       author = {{Jackson}, Alan P. and {Wyatt}, Mark C. and {Bonsor}, Amy and {Veras}, Dimitri},
        title = "{Debris froms giant impacts between planetary embryos at large orbital radii}",
      journal = {\mnras},
         year = 2014,
        month = jun,
       volume = {440},
       number = {4},
        pages = {3757-3777},
          doi = {10.1093/mnras/stu476},
archivePrefix = {arXiv},
       eprint = {1403.1888},
 primaryClass = {astro-ph.EP},
       adsurl = {https://ui.adsabs.harvard.edu/abs/2014MNRAS.440.3757J}
}

@article{Meng2014,
author = {Huan Y. A. Meng  and Kate Y. L. Su  and George H. Rieke  and David J. Stevenson  and Peter Plavchan  and Wiphu Rujopakarn  and Carey M. Lisse  and Saran Poshyachinda  and Daniel E. Reichart },
title = {Large impacts around a solar-analog star in the era of terrestrial planet formation},
journal = {Science},
volume = {345},
number = {6200},
pages = {1032-1035},
year = {2014},
doi = {10.1126/science.1255153},
URL = {https://www.science.org/doi/abs/10.1126/science.1255153},
eprint = {https://www.science.org/doi/pdf/10.1126/science.1255153}}

@ARTICLE{Paxton2011,
  author = {{Paxton}, B. and {Bildsten}, L. and {Dotter}, A. and {Herwig}, F. and {Lesaffre}, P. and {Timmes}, F.},
  title = {{Modules for Experiments in Stellar Astrophysics (MESA)}},
  journal = {\apjs},
  archivePrefix = {arXiv},
  eprint = {1009.1622},
  primaryClass = {astro-ph.SR},
  year = {2011},
  month = {jan},
  volume = {192},
  eid = {3},
  pages = {3},
  doi = {10.1088/0067-0049/192/1/3},
  adsurl = {https://ui.adsabs.harvard.edu/abs/2011ApJS..192....3P},
}

@ARTICLE{Paxton2013,
  author = {{Paxton}, B. and {Cantiello}, M. and {Arras}, P. and {Bildsten}, L. and {Brown}, E.~F. and {Dotter}, A. and {Mankovich}, C. and {Montgomery}, M.~H. and {Stello}, D. and {Timmes}, F.~X. and {Townsend}, R.},
  title = {{Modules for Experiments in Stellar Astrophysics (MESA): Planets, Oscillations, Rotation, and Massive Stars}},
  journal = {\apjs},
  archivePrefix = {arXiv},
  eprint = {1301.0319},
  primaryClass = {astro-ph.SR},
  year = {2013},
  month = {sep},
  volume = {208},
  eid = {4},
  pages = {4},
  doi = {10.1088/0067-0049/208/1/4},
  adsurl = {https://ui.adsabs.harvard.edu/abs/2013ApJS..208....4P},
}

@ARTICLE{Paxton2015,
  author = {{Paxton}, B. and {Marchant}, P. and {Schwab}, J. and {Bauer}, E.~B. and {Bildsten}, L. and {Cantiello}, M. and {Dessart}, L. and {Farmer}, R. and {Hu}, H. and {Langer}, N. and {Townsend}, R.~H.~D. and {Townsley}, D.~M. and {Timmes}, F.~X.},
  title = {{Modules for Experiments in Stellar Astrophysics (MESA): Binaries, Pulsations, and Explosions}},
  journal = {\apjs},
  archivePrefix = {arXiv},
  eprint = {1506.03146},
  primaryClass = {astro-ph.SR},
  year = {2015},
  month = {sep},
  volume = {220},
  eid = {15},
  pages = {15},
  doi = {10.1088/0067-0049/220/1/15},
  adsurl = {https://ui.adsabs.harvard.edu/abs/2015ApJS..220...15P},
}

@ARTICLE{Paxton2018,
  author = {{Paxton}, B. and {Schwab}, J. and {Bauer}, E.~B. and {Bildsten}, L. and {Blinnikov}, S. and {Duffell}, P. and {Farmer}, R. and {Goldberg}, J.~A. and {Marchant}, P. and {Sorokina}, E. and {Thoul}, A. and {Townsend}, R.~H.~D. and {Timmes}, F.~X.},
  title = {{Modules for Experiments in Stellar Astrophysics (MESA): Convective Boundaries, Element Diffusion, and Massive Star Explosions}},
  journal = {\apjs},
  archivePrefix = {arXiv},
  eprint = {1710.08424},
  primaryClass = {astro-ph.SR},
  year = {2018},
  month = {feb},
  volume = {234},
  eid = {34},
  pages = {34},
  doi = {10.3847/1538-4365/aaa5a8},
  adsurl = {https://ui.adsabs.harvard.edu/abs/2018ApJS..234...34P},
}

@ARTICLE{Paxton2019,
       author = {{Paxton}, Bill and {Smolec}, R. and {Schwab}, Josiah and {Gautschy}, A. and
         {Bildsten}, Lars and {Cantiello}, Matteo and {Dotter}, Aaron and
         {Farmer}, R. and {Goldberg}, Jared A. and {Jermyn}, Adam S. and
         {Kanbur}, S.~M. and {Marchant}, Pablo and {Thoul}, Anne and
         {Townsend}, Richard H.~D. and {Wolf}, William M. and {Zhang}, Michael and
         {Timmes}, F.~X.},
        title = "{Modules for Experiments in Stellar Astrophysics (MESA): Pulsating Variable Stars, Rotation, Convective Boundaries, and Energy Conservation}",
      journal = {\apjs},
         year = "2019",
        month = "Jul",
       volume = {243},
       number = {1},
          eid = {10},
        pages = {10},
          doi = {10.3847/1538-4365/ab2241},
archivePrefix = {arXiv},
       eprint = {1903.01426},
 primaryClass = {astro-ph.SR},
       adsurl = {https://ui.adsabs.harvard.edu/abs/2019ApJS..243...10P}
}

@ARTICLE{scargle,
       author = {{Scargle}, J.~D.},
        title = "{Studies in astronomical time series analysis. II. Statistical aspects of spectral analysis of unevenly spaced data.}",
      journal = {\apj},
         year = 1982,
        month = dec,
       volume = {263},
        pages = {835-853},
          doi = {10.1086/160554},
       adsurl = {https://ui.adsabs.harvard.edu/abs/1982ApJ...263..835S}
}

@ARTICLE{lomb,
       author = {{Lomb}, N.~R.},
        title = "{Least-Squares Frequency Analysis of Unequally Spaced Data}",
      journal = {\apss},
         year = 1976,
        month = feb,
       volume = {39},
       number = {2},
        pages = {447-462},
          doi = {10.1007/BF00648343},
       adsurl = {https://ui.adsabs.harvard.edu/abs/1976Ap&SS..39..447L}
}

@ARTICLE{2021ApJ...918...71R,
       author = {{Rieke}, G.~H. and {Su}, K.~Y.~L. and {Melis}, Carl and {G{\'a}sp{\'a}r}, Andr{\'a}s},
        title = "{Extreme Variability of the V488 Persei Debris Disk}",
      journal = {\apj},
         year = 2021,
        month = sep,
       volume = {918},
       number = {2},
          eid = {71},
        pages = {71},
          doi = {10.3847/1538-4357/ac0dc4},
archivePrefix = {arXiv},
       eprint = {2108.02901},
 primaryClass = {astro-ph.EP},
       adsurl = {https://ui.adsabs.harvard.edu/abs/2021ApJ...918...71R}
}

@ARTICLE{2020PSJ.....1...18C,
       author = {{Clement}, Matthew S. and {Kaib}, Nathan A. and {Chambers}, John E.},
        title = "{Embryo Formation with GPU Acceleration: Reevaluating the Initial Conditions for Terrestrial Accretion}",
      journal = {\psj},
         year = 2020,
        month = jun,
       volume = {1},
       number = {1},
          eid = {18},
        pages = {18},
          doi = {10.3847/PSJ/ab91aa},
archivePrefix = {arXiv},
       eprint = {2005.03668},
 primaryClass = {astro-ph.EP},
       adsurl = {https://ui.adsabs.harvard.edu/abs/2020PSJ.....1...18C}
}

@ARTICLE{Wyatt_Jackson_Review,
       author = {{Wyatt}, Mark C. and {Jackson}, Alan P.},
        title = "{Insights into Planet Formation from Debris Disks. II. Giant Impacts in Extrasolar Planetary Systems}",
      journal = {\ssr},
         year = 2016,
        month = dec,
       volume = {205},
       number = {1-4},
        pages = {231-265},
          doi = {10.1007/s11214-016-0248-1},
archivePrefix = {arXiv},
       eprint = {1603.04857},
 primaryClass = {astro-ph.EP},
       adsurl = {https://ui.adsabs.harvard.edu/abs/2016SSRv..205..231W}
}

@ARTICLE{Canup_GI,
       author = {{Canup}, Robin M.},
        title = "{Dynamics of Lunar Formation}",
      journal = {\araa},
         year = 2004,
        month = sep,
       volume = {42},
       number = {1},
        pages = {441-475},
          doi = {10.1146/annurev.astro.41.082201.113457},
       adsurl = {https://ui.adsabs.harvard.edu/abs/2004ARA&A..42..441C}
}

@ARTICLE{Wyatt_2008,
       author = {{Wyatt}, M.~C.},
        title = "{Evolution of debris disks.}",
      journal = {\araa},
         year = 2008,
        month = sep,
       volume = {46},
        pages = {339-383},
          doi = {10.1146/annurev.astro.45.051806.110525},
       adsurl = {https://ui.adsabs.harvard.edu/abs/2008ARA&A..46..339W}
}

@ARTICLE{Su_2019,
       author = {{Su}, Kate Y.~L. and {Jackson}, Alan P. and {G{\'a}sp{\'a}r}, Andr{\'a}s and {Rieke}, George H. and {Dong}, Ruobing and {Olofsson}, Johan and {Kennedy}, G.~M. and {Leinhardt}, Zo{\"e} M. and {Malhotra}, Renu and {Hammer}, Michael and {Meng}, Huan Y.~A. and {Rujopakarn}, W. and {Rodriguez}, Joseph E. and {Pepper}, Joshua and {Reichart}, D.~E. and {James}, David and {Stassun}, Keivan G.},
        title = "{Extreme Debris Disk Variability: Exploring the Diverse Outcomes of Large Asteroid Impacts During the Era of Terrestrial Planet Formation}",
      journal = {\aj},
         year = 2019,
        month = may,
       volume = {157},
       number = {5},
          eid = {202},
        pages = {202},
          doi = {10.3847/1538-3881/ab1260},
archivePrefix = {arXiv},
       eprint = {1903.10627},
 primaryClass = {astro-ph.EP},
       adsurl = {https://ui.adsabs.harvard.edu/abs/2019AJ....157..202S}
}

@ARTICLE{Vanderburg_WD1145,
       author = {{Vanderburg}, Andrew and {Johnson}, John Asher and {Rappaport}, Saul and {Bieryla}, Allyson and {Irwin}, Jonathan and {Lewis}, John Arban and {Kipping}, David and {Brown}, Warren R. and {Dufour}, Patrick and {Ciardi}, David R. and {Angus}, Ruth and {Schaefer}, Laura and {Latham}, David W. and {Charbonneau}, David and {Beichman}, Charles and {Eastman}, Jason and {McCrady}, Nate and {Wittenmyer}, Robert A. and {Wright}, Jason T.},
        title = "{A disintegrating minor planet transiting a white dwarf}",
      journal = {\nat},
         year = 2015,
        month = oct,
       volume = {526},
       number = {7574},
        pages = {546-549},
          doi = {10.1038/nature15527},
archivePrefix = {arXiv},
       eprint = {1510.06387},
 primaryClass = {astro-ph.EP},
       adsurl = {https://ui.adsabs.harvard.edu/abs/2015Natur.526..546V}
}

@ARTICLE{Oberst_16b,
       author = {{Oberst}, Thomas E. and {Rodriguez}, Joseph E. and {Col{\'o}n}, Knicole D. and {Angerhausen}, Daniel and {Bieryla}, Allyson and {Ngo}, Henry and {Stevens}, Daniel J. and {Stassun}, Keivan G. and {Gaudi}, B. Scott and {Pepper}, Joshua and {Penev}, Kaloyan and {Mawet}, Dimitri and {Latham}, David W. and {Heintz}, Tyler M. and {Osei}, Baffour W. and {Collins}, Karen A. and {Kielkopf}, John F. and {Visgaitis}, Tiffany and {Reed}, Phillip A. and {Escamilla}, Alejandra and {Yazdi}, Sormeh and {McLeod}, Kim K. and {Lunsford}, Leanne T. and {Spencer}, Michelle and {Joner}, Michael D. and {Gregorio}, Joao and {Gaillard}, Clement and {Matt}, Kyle and {Dumont}, Mary Thea and {Stephens}, Denise C. and {Cohen}, David H. and {Jensen}, Eric L.~N. and {Calchi Novati}, Sebastiano and {Bozza}, Valerio and {Labadie-Bartz}, Jonathan and {Siverd}, Robert J. and {Lund}, Michael B. and {Beatty}, Thomas G. and {Eastman}, Jason D. and {Penny}, Matthew T. and {Manner}, Mark and {Zambelli}, Roberto and {Fulton}, Benjamin J. and {Stockdale}, Christopher and {DePoy}, D.~L. and {Marshall}, Jennifer L. and {Pogge}, Richard W. and {Gould}, Andrew and {Trueblood}, Mark and {Trueblood}, Patricia},
        title = "{KELT-16b: A Highly Irradiated, Ultra-short Period Hot Jupiter Nearing Tidal Disruption}",
      journal = {\aj},
         year = 2017,
        month = mar,
       volume = {153},
       number = {3},
          eid = {97},
        pages = {97},
          doi = {10.3847/1538-3881/153/3/97},
archivePrefix = {arXiv},
       eprint = {1608.00618},
 primaryClass = {astro-ph.EP},
       adsurl = {https://ui.adsabs.harvard.edu/abs/2017AJ....153...97O}
}

@ARTICLE{Li2010_gas_giant,
       author = {{Li}, Shu-Lin and {Miller}, N. and {Lin}, Douglas N.~C. and {Fortney}, Jonathan J.},
        title = "{WASP-12b as a prolate, inflated and disrupting planet from tidal dissipation}",
      journal = {\nat},
         year = 2010,
        month = feb,
       volume = {463},
       number = {7284},
        pages = {1054-1056},
          doi = {10.1038/nature08715},
archivePrefix = {arXiv},
       eprint = {1002.4608},
 primaryClass = {astro-ph.EP},
       adsurl = {https://ui.adsabs.harvard.edu/abs/2010Natur.463.1054L}
}
\bibliographystyle{aasjournal}

\newpage
\clearpage 

\appendix

\section{Cluster Membership Probability}\label{ap:apendixA}

\begin{figure*}[h!]
    \centering
    \includegraphics[width=0.98\linewidth]{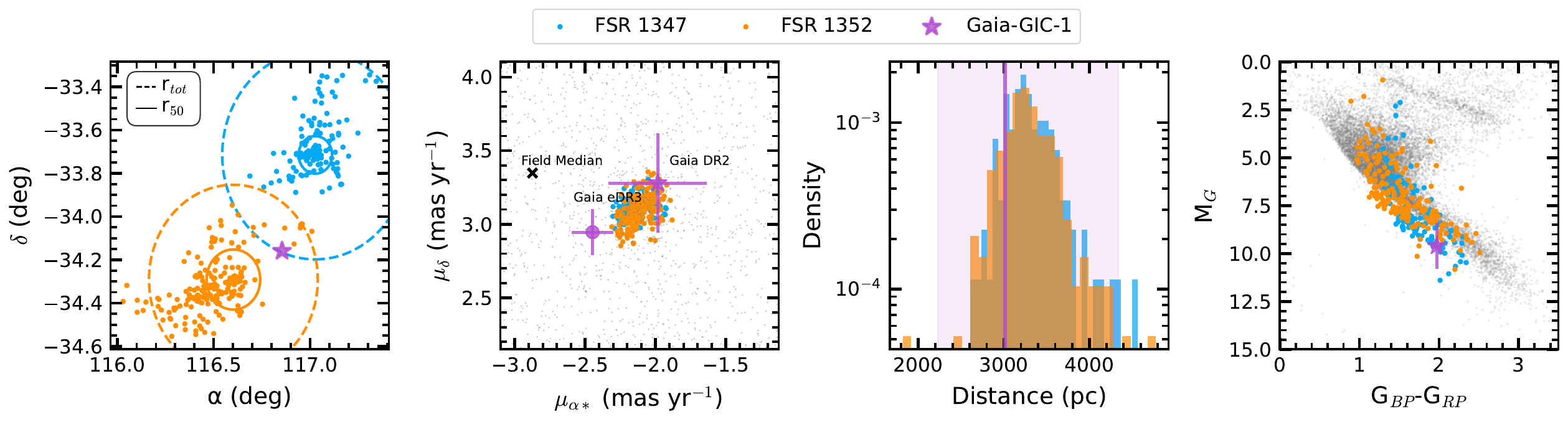}
    \caption{Astrometric and photometric properties of FSR 1347, FSR 1352, and Gaia-GIC-1 (magenta star). (\textbf{Left}): Sky distribution of the two open clusters, where the solid lines represent the 50\% cluster membership and dashed lines the total cluster radius \citep{Hunt_2024}. (\textbf{Center Left}) Proper motions of the identified clusters, including the reported 2D proper motions, are reported in both Gaia DR2 and Gaia DR3. For reference, we also include the field median marked. (\textbf{Center Right}) Probability distribution of the reported cluster member distances and Gaia-GIC-1 from \citet{BailerJonesDR3}. (\textbf{Right}) Extinction corrected Gaia DR3 color-magnitude diagram including the open cluster members and the position of Gaia-GIC-1.}
    \label{fig:gaia_diagnostics}
\end{figure*}

We performed a Gaussian Mixture Model (GMM) analysis to assess the cluster membership probability of Gaia-GIC-1 of two identified open clusters within $<$30-arcminutes. We constructed a training set consisting of flagged members of FSR 1347 and FSR 1352 from Gaia DR3 \citep{Hunt_2024}, along with field stars within the same sky footprint. We initialized our GMM on K=3 clusters, and trained it on the three-dimensional astrometry available from Gaia DR3, including parallax and the proper motions ($\mu_{\alpha} \text{cos}(\delta)$, $\mu_{\delta}$, $\omega$). To account for measurement uncertainties, we employed a Monte Carlo approach, drawing 1000 realizations from the astrometric parameters for Gaia-GIC-1 distributions centered on the measured values with widths of 2$\sigma$ of the reported Gaia DR3 uncertainties. For each realization, we computed the posterior probability of membership of each component. We found a median membership probability of P(FSR 1352)=0.6, compared to P(Field)=0.38 and P(FSR 1347)$\approx$0, suggesting marginal evidence that Gaia-GIC-1 might be a member of FSR 1352, though we caution that this should be done more concretely if radial velocities are obtained for Gaia-GIC-1 and both clusters. We present our diagnostic plots in Figure \ref{fig:gaia_diagnostics}.

\end{document}